\documentclass[%
reprint,
superscriptaddress,
 amsmath,amssymb,
prl,
]{revtex4-2}

\pdfoutput=1 
\usepackage{graphicx}
\usepackage{dcolumn}
\usepackage{bm}
\usepackage{physics}
\usepackage[english]{babel}

\newcommand{\Asym}{\text{Asym}}
\newcommand{\G}{\mathbb{G}}
\newcommand{\U}{\mathbb{U}}
\newcommand{\Id}{\mathbb{I}}
\newcommand\numthis{\stepcounter{equation}\tag{\theequation}}

\begin{document}

\preprint{APS/123-QED}

\title{Maximum Electromagnetic Local Density of States via Material Structuring}

\author{Pengning Chao}
\thanks{These two authors contributed equally}
\affiliation{Department of Electrical and Computer Engineering, Princeton University, Princeton, New Jersey 08544, USA}
\altaffiliation[email: ]{pengning@princeton.edu}
\author{Rodrick Kuate Defo}%
\thanks{These two authors contributed equally}
\affiliation{Department of Electrical and Computer Engineering, Princeton University, Princeton, New Jersey 08544, USA}
\author{Sean Molesky}
\affiliation{Department of Engineering Physics, Polytechnique Montréal, Montréal, Québec H3T 1J4, Canada}
\author{Alejandro Rodriguez}
\affiliation{Department of Electrical and Computer Engineering, Princeton University, Princeton, New Jersey 08544, USA}
\date{\today}

\begin{abstract}
The electromagnetic local density of states (LDOS) is crucial to many aspects of photonics engineering, from enhancing emission of photon sources to radiative heat transfer and photovoltaics. We present a framework for evaluating upper bounds on the LDOS in structured media that can handle arbitrary bandwidths and accounts for critical wave scattering effects. The bounds are solely determined by the bandwidth, material susceptibility, and device footprint, with no assumptions on geometry.
We derive an analytical expression for the maximum LDOS consistent with the conservation of energy across the entire design domain, which upon benchmarking with topology-optimized structures is shown to be nearly tight for large devices. Novel scaling laws for maximum LDOS enhancement are found: the bounds saturate to a finite value with increasing susceptibility and scale as the quartic root of the bandwidth for semi-infinite structures made of lossy materials, with direct implications on material selection and design applications.
\end{abstract}

\maketitle


\section{Introduction}

The electromagnetic local density of states (LDOS), a measure of local
response---the electric field from a dipolar current source---at a
given point in space~\cite{novotny_principles_2012}, plays a central role in many nanophotonic
phenomena and applications, including spontaneous~\cite{purcell_absorption_1969} and stimulated~\cite{francs_optical_2010}
emission, quantum information~\cite{barnes_classical_2020}, surface enhanced Raman scattering~\cite{michon_limits_2019,gaponenko_strong_2021}, photovoltaics~\cite{wang_optimization_2014}, radiative
heat transfer~\cite{cuevas_radiative_2018}, non-linear frequency conversion~\cite{lin_cavity-enhanced_2016}, scintillation~\cite{roques-carmes_framework_2022}, to name a
few. Enhancing the LDOS by nanostructuring bulk media, a persistent theme
in photonics design, is often achieved through the creation of optical
resonances~\cite{vahala_optical_2003}: a cavity or resonator supporting
a single mode of quality factor $Q$ and mode volume $V$ enhances the LDOS
in proportion to the Purcell factor $Q/V$. Many classic design schemes
rely on maximizing $Q$ (ring
resonators~\cite{armani_ultra-high-q_2003}), minimizing $V$ (plasmonic nanocavities~\cite{baumberg_extreme_2019,chikkaraddy_single-molecule_2016},
bow-tie antennae~\cite{kinkhabwala_large_2009}), or a combination thereof (photonic crystal
cavities~\cite{joannopoulos_photonic_2008}, metal-dielectric hybrid
structures~\cite{karamlou_metal-dielectric_2018}). Going beyond the single mode picture, multi-mode and
dispersion-based effects such as exceptional
points~\cite{pick_general_2017,lin_enhanced_2016} and slow light
devices~\cite{dewhurst_slow-light-enhanced_2010} have also been
explored. Recently, complex devices obtained by application of large-scale structural optimization appear to combine several confinement mechanisms~\cite{wang_maximizing_2018,albrechtsen_nanometer-scale_2021}, providing further improvements on device performance.

Given its relevance in optics and the ever increasing structural
freedom brought on by advances in nanofabrication~\cite{molesky_inverse_2018}, interest in
assesing limits on achievable LDOS enhancement has grown over the past
few decades~\cite{chao_physical_2022}. Spectral sum
rules~\cite{barnett_sum_1996,scheel_sum_2008,sanders_analysis_2018}
pin the frequency-integrated LDOS of any system but are of limited
utility in practical settings involving finite and often narrow source
bandwidths. Specialized results pertaining to achievable mode quality
factors~\cite{raman_upper_2013} or mode volumes in cavity settings~\cite{zhao_minimum_2020}
have proven useful in many applications but are not sufficiently
general to account for multi-mode effects. Passivity requirements
based on achievable material response were recently used to bound both
single-frequency~\cite{miller_fundamental_2016} and
finite-bandwidth~\cite{shim_fundamental_2019} LDOS. However, passivity
alone cannot fully capture the wave nature of light as constrained by
Maxwell's equations, e.g., the necessity of phase matching to achieve
resonance, leading to loose limits when contributions other than those coming from evanescent fields become relevant. 

In this article, we generalize and lift several limitations imposed by
these prior approaches to provide expressions and predictions for the
largest spectral-integrated LDOS that may be achieved in a structured
medium. The derived bounds incorporate wave and material constraints
imposed by Maxwell's equations over any desired length-scale and are
geometry agnostic: besides specifying the material susceptibility
$\chi$, a frequency window of interest, and a bounding domain that the
structured medium must reside within, no further assumptions are made
on device topology. We consider sources both enclosed within and
external to devices, obtaining bounds that generally come within an
order of magnitude of complicated structures discovered through
inverse methods~\cite{liang_formulation_2013}. Furthermore, we
find that the mere requirement of energy conservation is
sufficient to place tight limits on large devices; we exploit this fact to derive an analytical upper bound (\ref{eq:theta_dual}) along with integral expressions and asymptotic analysis for the particular scenario of a source above a semi-infinite, structured slab. The bounds show
saturation to a finite value as $\chi \rightarrow \infty$ and varying power-law scalings with respect to source bandwidth $\Delta \omega$, with the maximum LDOS transitioning from a scaling $\propto \Delta \omega^{-1}$ to one $\propto \Delta \omega^{-1/4}$ as material absorption begins to limit the net enhancement that any one mode may contribute. 

\section{Problem Formulation}
Working in dimensionless units of $\epsilon_0=\mu_0=1$, and
considering only nonmagnetic materials, the partial LDOS at frequency
$\omega$ and position $\vb{x}'$ along the direction $\hat{\vb{e}}$ can
be shown to be, by Poynting's theorem~\cite{jackson_classical_1999}, directly proportional to the
average power $\rho(\omega)$ emitted by a harmonic dipole source
$\vb{J}(\vb{r})e^{-i\omega t} = e^{-i\omega t} \delta(\vb{x}-\vb{x'})
\hat{\vb{e}}$ at the same location,
\begin{equation}
    \rho(\omega) \equiv -\frac{1}{2} \Re{\int \vb{J}^*(\vb{r}) \cdot \vb{E}(\vb{r}) \,\dd\vb{r} }
\end{equation}
where the electric field generated by the $\vb{J}$ solves Maxwell's equations,
\begin{equation}
    \curl \curl \vb{E}(\vb{r}) - \omega^2 \epsilon(\vb{r}) \vb{E}(\vb{r}) = i\omega\vb{J}(\vb{r}) .
\end{equation}

Since real sources emit light over a finite bandwidth (e.g., the Planck
spectrum of heated bodies or fluorescence/spontaneous emission
linewidth of atoms~\cite{hindmarsh_atomic_2014}), we instead consider the more natural choice
of a frequency average of $\rho(\omega)$ over a Lorentzian lineshape
centered at $\omega_0$ with bandwidth $\Delta\omega_{src} = \omega_0/2
Q_{src}$ (corresponding source ``quality factor'' $Q_{src}$):
\begin{subequations}
\begin{align}
    \langle \rho \rangle_{\omega_0, Q_{src}} &= \int_{-\infty}^\infty \frac{\Delta\omega_{src}/\pi}{(\omega-\omega_0)^2 + \Delta\omega_{src}^2} \rho(\omega) \,\dd\omega, 
\label{eq:freq_avg}
\end{align}
\end{subequations}
As pointed out in ~\cite{liang_formulation_2013}, such a modified figure of merit not only
captures the spectral lineshape of many practical sources~\cite{hindmarsh_atomic_2014}, but
also offers a computational and conceptual advantage: instead of
evaluating the frequency integral in (\ref{eq:freq_avg}) directly, one
can instead carry out a complex-$\omega$ contour integral over the
upper half plane and exploit causality to convert, using the residue
theorem, the spectral average $\langle \rho \rangle$ to the single
complex-frequency $\rho(\tilde{\omega})$ and an electrostatic (zero-frequency) contribution:
\begin{equation}
    \langle \rho \rangle_{\omega_0, Q_{src}} = \rho\bigg(\tilde{\omega}\equiv \omega_0 + \frac{\omega_0}{2Q_{src}} i \bigg) + \frac{4Q_{src}}{\pi(4Q_{src}^2 + 1) \omega_0} \alpha.
    \label{eq:cplx_freq_rho}
\end{equation}
Here $\alpha=\frac{1}{2} \Re{ \vb{p}_0 \cdot \vb{E}_0}$, where $\vb{p}_0$ is a unit amplitude electrostatic dipole and $\vb{E}_0$ the field it generates; this electrostatic contribution leads to an all-frequency integrated sum rule in the wide bandwidth limit~\cite{sanders_analysis_2018,shim_fundamental_2019}. In this paper we will be focusing on the practically important case of moderate or narrow source bandwidth with $Q_{src} \gg 1$ and thus will neglect the electrostatic term in (\ref{eq:cplx_freq_rho}) going forward.

Thus, the structural design problem for maximizing bandwidth-averaged
LDOS can be formulated as
\begin{subequations}
\begin{align}
    \max_{\epsilon(\vb{r}; \tilde{\omega})} \quad \rho(\tilde{\omega}) = -\frac{1}{2}\Re{\int \vb{J}^*(\vb{r};\tilde{\omega}) \cdot \vb{E}(\vb{r}; \tilde{\omega}) \,\dd\vb{r} }
    \label{eq:TO_objective}
\end{align}
given the constraints
\begin{align}
    &\curl \curl \vb{E}(\vb{r}; \tilde{\omega}) - \tilde{\omega}^2 \epsilon(\vb{r}; \tilde{\omega}) \vb{E}(\vb{r}; \tilde{\omega}) = i\tilde{\omega}\vb{J}(\vb{r}; \tilde{\omega}) \\
    &\epsilon(\vb{r}; \tilde{\omega}) = 
    \begin{cases}
    1 \text{ or } 1+\chi(\tilde{\omega}) & \vb{r} \in V \\
    1 &  \vb{r} \notin V
    \end{cases}
\label{eq:structural_opt} 
\end{align}
\end{subequations}
where $V$ is a pre-specified design region that the structure resides
within and $\chi(\tilde{\omega})$ the Fourier transform of the bulk material
susceptibility evaluated at complex frequency $\tilde{\omega}$. In later expressions we may suppress the $\tilde{\omega}$ dependence of time-harmonic variables for notation simplicity.

\begin{figure*}
\includegraphics[width=\linewidth]{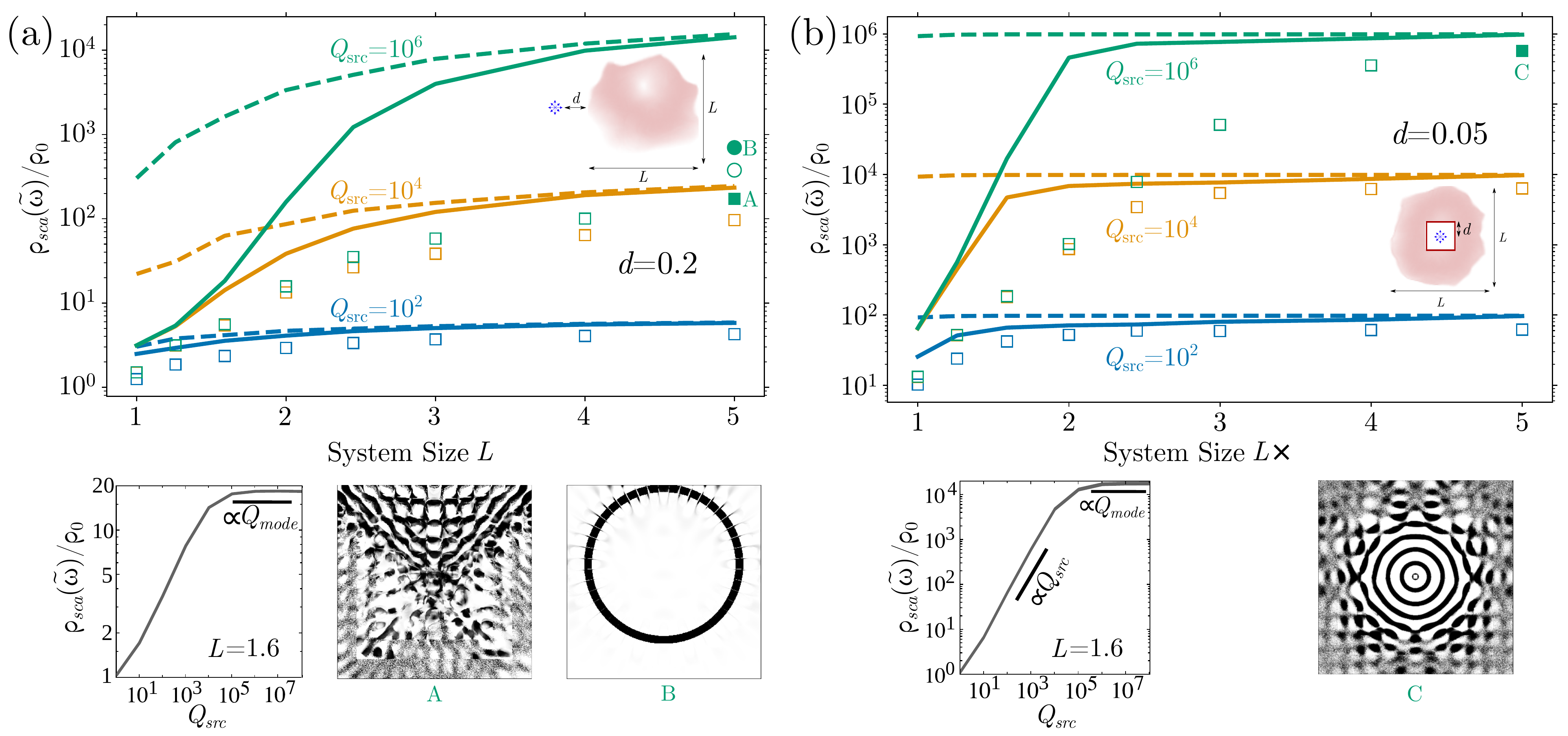}
\caption{\label{fig:FDFD_aggregate} Bounds and inverse designs with
  TM polarization, $\chi=4$, $Q_{src}\in \{10^2, 10^4, 10^6\}$, and finite design domain size $L$ for (a) dipole on the side of
  a square design region and (b) dipole in the center of a square
  design region. Solid lines are bounds that are converged with
  respect to increasing the number of constraint subregions
  $V_k$. Dotted lines are bounds with just the global constraint
  $V_k=V$. Squares are topology optimized structures from random
  initializations. The hollow circle is a ring resonator with inner
  and outer radii approximately 1.85 and 2.05 respectively, with the exact resonant radius requiring accuracy up to 8 significant figures to achieve the plotted enhancement (SI); the filled
  circle is a topology optimized structure starting from that ring
  resonator. LDOS enhancement as a function of $Q_{src}$ for fixed $L=1.6$ is shown beneath the main figures, along with designs corresponding to filled shapes.}
\end{figure*}

Due to the high dimensionality of the
structural degrees of freedom $\epsilon(\vb{r})$ and the non-convex dependence of the field $\vb{E}(\vb{r})$ on $\epsilon(\vb{r})$, it is generally not possible to solve for the global optimum of (\ref{eq:structural_opt})~\cite{christiansen_inverse_2021}. However, 
it is possible to exploit an alternative parametrization of the
problem in which the polarization density
$\vb{P}(\vb{r})=(\epsilon(\vb{r})-1)\vb{E}(\vb{r})$ rather than
$\epsilon(\vb{r})$ serve as optimizations degrees of freedom, to
obtain bounds on $\rho(\tilde{\omega})$ applicable to \emph{any possible
structure}, given only the choices of design region $V$ and material
susceptibility $\chi$~\cite{chao_physical_2022}. In this context, it becomes convenient to
decompose the total field into the vacuum field emitted by the
source and scattered field emitted by the polarization currents:
\begin{align*}
    &\vb{E}(\vb{r}) = \vb{E}_{vac}(\vb{r}) + \vb{E}_{sca}(\vb{r}) \\
    &= \frac{i}{\tilde{\omega}} \int \G(\vb{r},\vb{r'}) \cdot \vb{J}(\vb{r'}) \,\dd\vb{r'} + \int \G(\vb{r},\vb{r'}) \cdot \vb{P}(\vb{r'}) \,\dd\vb{r'} \numthis
\end{align*}
with $\G(\vb{r},\vb{r'})$ denoting the vacuum dyadic Green's function, 
the solution to
\begin{equation}
    \curl \curl \G - \tilde{\omega}^2 \G = \tilde{\omega}^2 \Id .
\end{equation}
Note that our definition of the Green's function has an extra global
factor of $\tilde{\omega}^2$ compared to the standard definition.
Similarly, the net power extracted from the dipole source can be
decomposed into constant (structure-independent) vacuum and
scattered-field contributions:
\begin{subequations}
\begin{equation}
    \rho = \rho_{vac} + \rho_{sca}(\vb{P}) ,
\end{equation}
\begin{equation}
    \rho_{vac} = -\frac{1}{2}\Re{\frac{i}{\tilde{\omega}} \iint \vb{J}^*(\vb{r}) \cdot \G(\vb{r},\vb{r'}) \cdot \vb{J}(\vb{r'}) \,\dd\vb{r'} \,\dd\vb{r} } ,
\end{equation}
\begin{align*}
    \rho_{sca}(\vb{P}) &= -\frac{1}{2} \Re{ \iint \vb{J}^*(\vb{r}) \cdot \G(\vb{r},\vb{r'}) \cdot \vb{P}(\vb{r'}) \,\dd\vb{r'} \,\dd\vb{r}} \\
    &= \frac{1}{2} \Im{ \tilde{\omega} \int \vb{E}_{vac}(\vb{r'}) \cdot \vb{P}(\vb{r'}) \,\dd\vb{r'} } ,\numthis
    \label{eq:rho_sca}
\end{align*}
\end{subequations}
where in the second line of (\ref{eq:rho_sca}) we made use of the fact
that $\vb{J}^* = \vb{J}$ (the global phase of dipole source is
irrelevant) and the reciprocity relation $\G(\vb{r},\vb{r'}) =
\G^T(\vb{r'},\vb{r})$. They key to formulating a shape-independent
bound on $\rho$ is to forgo the need for structural information by
relaxing the requirement that $\vb{P}$ satisfy Maxwell's equations
everywhere, imposing instead a smaller subset of wave
constraints~\cite{chao_physical_2022}. Such constraints can be indeed be derived from
Maxwell's equations through a complex frequency generalization of the
time-harmonic Poynting's theorem (see~\cite{chao_physical_2022} and SI), giving
\begin{align*}
    &\int_{V_k} \vb{E}_{vac}(\vb{r}) \cdot \vb{P}(\vb{r}) \,\dd\vb{r} = \int_V \vb{P}^*(\vb{r}) \cdot \int_{V_k} \U(\vb{r},\vb{r'}) \cdot \vb{P}(\vb{r'}) \,\dd\vb{r'} \,\dd\vb{r} ,\\
    &\U \equiv \chi^{*-1} \delta(\vb{r}-\vb{r'}) - \G^*(\vb{r},\vb{r'}) \numthis
    \label{eq:constraints}
\end{align*}
where $V_k \subseteq V$ is any spatial region within the design domain
$V$ and we have defined the composite operator $\U$. Notably,
(\ref{eq:constraints}) can be interpreted as a statement of the
conservation of energy over each spatial region, requiring only
specification of the allotted design footprint $V$ and available
susceptibility $\chi$. Note also that geometric information is
contained only implicitly in $\vb{P}$, with material properties
specified by the known complex scalar $\chi$. The resulting
optimization problem for the LDOS can be written as
\begin{subequations}
\begin{equation}
    \max_{\vb{P}(\vb{r}; \tilde{\omega})} \quad \rho_{sca} =
    -\frac{1}{2} \Im{ \tilde{\omega} \bra{\vb{E}_{vac}^*}\ket{\vb{P}} }
    \label{eq:primal_objective}
\end{equation}
such that $\forall V_k \subseteq V$
\begin{multline}
     \Re\text{,}\Im \big\{ \bra{\vb{E}_{vac}}\Id_{V_k}\ket{\vb{P}} - \bra{\vb{P}} \U\Id_{V_k}\ket{\vb{P}} \big\}= 0
    \label{eq:primal_constraint}
\end{multline}
\label{eq:primal_problem}%
\end{subequations}
where $\Id_{V_k}$ is the projection operator into the region
$V_k$. Above, we employed Dirac bra-ket notation for brevity, with
$\ket{\vb{a}}$ denoting the vector field $\vb{a}(\vb{r})$ over the
design domain $V$, $\bra{\vb{a}}\ket{\vb{b}}=\int_V \vb{a}^*(\vb{r})
\cdot \vb{b}(\vb{r}) \,\dd\vb{r}$ is the conjugated inner product, and
operator action $\mathbb{A} \ket{\vb{a}} = \int_V
\mathbb{A}(\vb{r},\vb{r'}) \cdot \vb{a}(\vb{r'}) \,\dd\vb{r'}$.
(\ref{eq:primal_problem}) is a quadratically constrained linear
program for $\vb{P}$. While direct optimization is still
not guaranteed to find a global optimum due to the non-convexity of certain constraints in
(\ref{eq:primal_constraint}), a bound on the problem can be computed
efficiently via the Lagrange dual
function~\cite{angeris_heuristic_2021,chao_physical_2022,boyd_convex_2004}, hereby referred to as the dual bound.

In principle, the maximum LDOS achievable in a structured
medium is known to grow indefinitively in the single-frequency limit
of $\Delta\omega_{src}\to 0$. For instance, in lossless media, the whispering
gallery modes of ring resonators and defect modes of photonic
crystals exhibit exponential scaling in $Q$ with increasing system
size~\cite{marcatili_bends_1969,joannopoulos_photonic_2008}, while
their mode volumes either increase polynomially or remain constant,
respectively, leading to exponential growth in the Purcell
factor. Geometries with sharp tips can also give rise to field
singularities and hence vanishing mode
volumes~\cite{budaev_electromagnetic_2007}. However, in practice,
finite source bandwidths, device footprint, and material losses limit
the utility of diverging lifetimes, while fabrication imperfections
and atomic-scale effects preclude realization of arbitrarily small
features. As will be seen below, the imposition of finite $\Delta
\omega_{src}$ and a minimum vacuum gap $d$ between source and medium
regularizes such divergences, paving the way for
investigations of LDOS growth characteristics in realistic settings. 
We also note that for finite $\Delta \omega_{src}$ the vacuum LDOS contribution
$\rho_{vac}$ diverges due to contributions from high frequencies. This is an
artifact of the Lorentzian lineshape chosen and has little bearing on practical applications; the results presented will thus focus on the structural contribution $\rho_{sca}$ which does remain finite. 

To make explicit the scale invariance of Maxwell's equations, all lengths are given in
units of the center wavelength $\lambda_0 = 2\pi c/\omega_0$ and
$\rho_{sca}$ normalized by the single-frequency vacuum
dipole radiation $\rho_0 \equiv \rho_{vac}(\omega_0)$.

\begin{figure*}
\includegraphics[width=\linewidth]{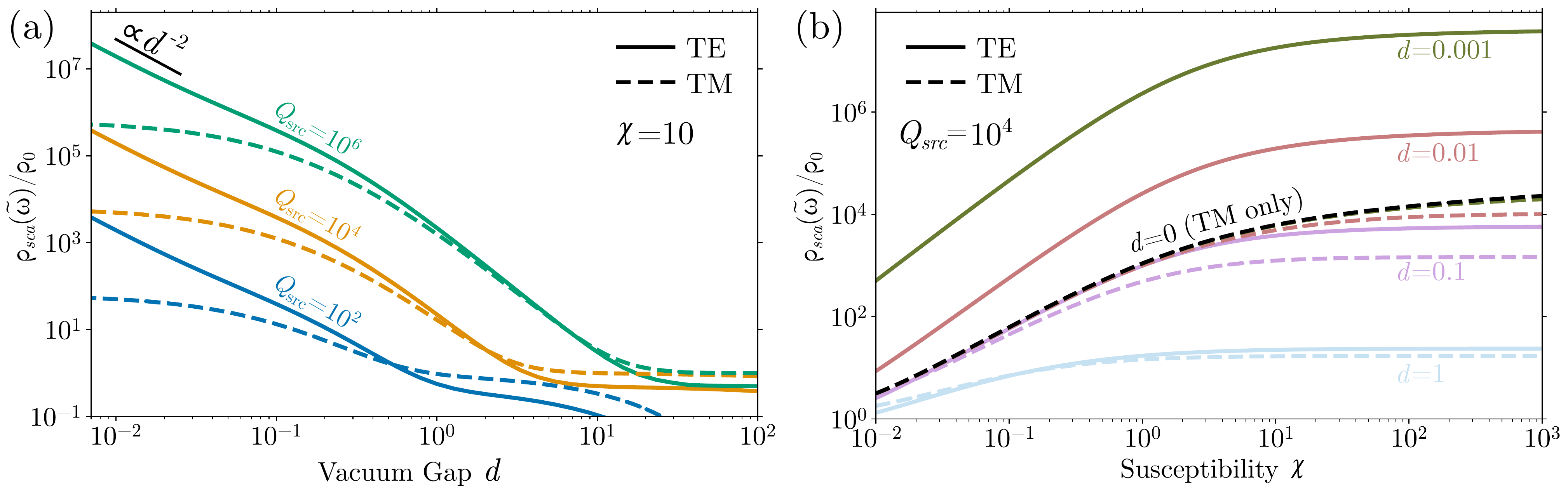}
\caption{\label{fig:halfspace_lossless} Maximum LDOS enhancement near a half-space design region as a function of (a) separation distance and (b) real material susceptibility $\chi$ for lossless dielectrics. While (b) plots bounds for lossless dielectrics, the material independent limit as $\chi \rightarrow \infty$ is a bound for general lossy $\chi$ as well (SI). }
\end{figure*}

\section{Numerical Results}

The dual bound can be obtained numerically for arbitrary domains via a
suitable representation of the Green's function and Maxwell
operator. We do so for two standard settings: an external--dipole
configuration where the dipole is adjacent to the structure, relevant
for instance to gratings and solid-state defect couplers~\cite{chakravarthi_inverse-designed_2020,wambold_adjoint-optimized_2021}; and
an enclosed--dipole configuration in which the dipole is surrounded by
the structure, relevant to photonic crystal cavities~\cite{joannopoulos_photonic_2008}, bullseye
gratings~\cite{li_efficient_2015}, and bowtie antennas~\cite{kinkhabwala_large_2009}. For computational convenience,
all calculations are performed in 2d for both the out-of-plane
(scalar) TM and in-plane (vectorial) TE electric-field
polarizations. The finite-difference frequency-domain method is used
to represent all relevant fields and operators
(SI). Progressively tighter bounds were obtained by gradually
increasing the number of constraint sub-regions $V_k$ down to the
computational pixel level. Inverse designs were also obtained to
compare against the bounds, following the topology optimization (TO)
approach detailed in~\cite{liang_formulation_2013}.

Results pertaining to maximum LDOS for both external and enclosed
configurations are shown in Figure~\ref{fig:FDFD_aggregate} as a
function of the design footprint. Since the vacuum LDOS diverges when
integrated against a Lorentzian
lineshape~\cite{barnett_sum_1996,liang_formulation_2013,shim_fundamental_2019},
this and subsequent results pertain only to the structural
contribution $\rho_{sca}$, defining the enhancement factor in relation
to the single-frequency vacuum emission $\rho_0 \equiv
\rho_{vac}(\omega_0)$. In both settings the medium is assumed to be
lossless, $\Im\chi=0$. For a fixed $Q_{src}$, the bounds are
seen to grow exponentially before saturating with increasing domain
size $L$. Conversely, for fixed $L$ the bounds grow linearly before
saturating with increasing $Q_{src}$. Both observations are consistent
with a resonant enhancement mechanism: since the
finite vacuum gap $d$ precludes field divergences in the vicinity of
the dipole, thereby imposing an upper bound on its coupling to any
mode, one would expect improvements in the Purcell factor to be mainly
driven by growth in the modal lifetimes. Supposing a system with
response dominated by a single mode of quality factor $Q_{mode} =
\omega_0/2\Delta\omega_{mode}$, (\ref{eq:freq_avg}) yields a
frequency-integrated LDOS that scales as
\begin{equation}
    \int_{-\infty}^\infty \frac{\Delta\omega_{src}/\pi}{\omega^2+\Delta\omega_{src}^2} \frac{\Delta\omega_{mode}/\pi}{\omega^2+\Delta\omega_{mode}^2} \,\dd\omega = \frac{Q_{src} Q_{mode}}{2\pi\omega_0(Q_{src}+Q_{mode})} .
    \label{eq:single_mode_avg}
\end{equation}
Thus, this modal picture predicts linear growth $\propto Q_{src}$ in
(\ref{eq:freq_avg}) so long as the system supports a sufficiently
long-lived mode, $Q_{mode} \gg Q_{src}$, eventually saturating to a
value $\propto Q_{mode}$ whenever $Q_{mode} \ll Q_{src}$. In the
absence of material dissipation, the highest achievable $Q_{mode}$ is
constrained solely by radiative losses and, as confirmed by our bound
calculations, known to scale at least exponentially with $L$ in
ring-resonator and photonic-crystal cavities. Comparing our bounds
(solid lines) against inverse designs (squares), one observes
remarkable alignment, with bounds and designs often coming within an
order of magnitude of each other. The biggest performance gaps occur
at high $Q_{src}$ in the external configuration and can at least be
partially attributed to TO getting trapped in local optima that,
regardless of exhaustive initial conditions, underperform simple ring resonator geometries. Even starting with ring resonators
as initial seeds, TO was only able to make modest improvements. In
contrast, Bragg--onion type structures discovered by TO in the
cavity setting are seen to tightly approach
corresponding bounds. Comparing the high degree of precision needed to specify the dimensions of the ring resonator against the relative robustness of photonic bandgap confinement~\cite{rodriguez_disorder-immune_2005}, we hypothesize that attaining high-performing designs for the external configuration at narrow bandwidths requires resonances based on sensitive interference cancellation. This leads to an ill-behaved optimization landscape with many subpar local optima to get stuck in, with ring resonators as an example of a class of high-performing designs not easily discoverable via brute-force optimization. 

Finally, the observed dependence of the dual bound on the number and
size of subregion constraints (\ref{eq:primal_constraint}) reveals two
important features: first, the necessity of imposing many
subwavelength constraint regions (SI) for the bounds to exhibit
saturation as $Q_{src} \to \infty$; second, the fact that a single,
global energy-conservation constraint appears sufficient to produce
tight bounds when either device sizes or source bandwidths become
sufficiently large. This suggests that subregion constraints are
necessary to adequately ``resolve'' wave effects, including
phase-matching restrictions, that limit mode confinement in finite
systems; conversely, when system sizes no longer limit the highest
achievable $Q_{mode}$ in relation to $Q_{src}$, the latter and not the
former becomes a bottleneck for further enhancements.

\section{Large-size Semi-analytics}
We now exploit the remarkable success of global energy conservation in describing large system behavior to obtain
a semi-analytical bound on the maximum LDOS above a semi-infinite
structure, i.e., the $L\rightarrow\infty$ limit of the external--dipole
configuration depicted in Figure~\ref{fig:FDFD_aggregate}a. As shown
in SI, the solution of the dual problem under both resistive and
reactive energy conservation constraints (\ref{eq:primal_constraint})
can be mapped to a family of solutions involving a unitary phasor
parametrized by the single constraint angle $\theta$, leading to the
modified problem:
\begin{subequations}
\begin{equation}
  \max_{\vb{P}(\vb{r}; \tilde{\omega})}  \quad \rho_{sca} = -\frac{1}{2} \Im{ \tilde{\omega} \bra{\vb{E}_{vac}^*}\ket{\vb{P}} }
\label{eq:theta_problem}  
\end{equation}
such that for $\mathbf{P} \in V$,
\begin{equation}
\Im{e^{i\theta} \bra{\vb{E}_{vac}}\ket{\vb{P}}} - \bra{\vb{P}} \Asym(e^{i\theta}\U)\ket{\vb{P}}) = 0
\label{eq:theta_constraint}
\end{equation}
\end{subequations}
where $\Asym(\mathbb{A}) = (\mathbb{A}-\mathbb{A}^\dagger)/2i$ is the Hermitian anti-symmetric component of the operator $\mathcal{A}$.
Owing to its simplicity, the solution to this dual bound for any given
$\theta$ can be written explicitly as
\begin{multline}
    \rho_{sca} \leq \frac{1}{4} \mathrm{Re}\Big\{\Big(-\tilde{\omega}^* e^{i\theta}
      \ket{\vb{E}^*_{vac}}+ |\tilde{\omega}| \ket{\vb{E}_{vac}}\Big)^\dagger \\
      \Asym(e^{i\theta}\U)^{-1} \ket{\vb{E}_{vac}}\Big\},
    \label{eq:theta_dual}
\end{multline}
which notably, depends only on the analytically known quantities
$\vb{E}_{vac}$, $\chi$, and $\U$. The tightest dual bound consistent
with (\ref{eq:theta_dual}) can thus be obtained numerically by
carrying out an additional optimization over $\theta$, restricted so
that $\Asym(e^{i\theta} \U)$ is positive definite (SI). Further analytical insight can be gleaned by expanding the
operators and fields in a spectral basis conforming to the symmetry of
the design domain: for a half-space enclosure, the natural choice is a
Fourier basis $e^{ik_\parallel x_\parallel}$ parametrized by the
wavevector $k_\parallel$ parallel to the half-space surface. Carrying
out this expansion for the incident field $\ket{\vb{E}_{vac}}$ yields,
\begin{subequations}
\begin{equation}
    \ket{\vb{E}_{vac}} = -\frac{\tilde{\omega}}{2\sqrt{2\pi}}
    \int_{-\infty}^\infty \frac{e^{ik_\parallel
        x}}{\sqrt{2\pi}}\frac{1}{k_\perp} e^{ik_\perp x_\perp} e^{ik_\perp d} \dd k_\parallel
\end{equation}
\end{subequations}
where $k_\perp = \sqrt{\tilde{\omega}^2-k_\parallel^2}$ and
$\Im(k_\perp)\geq0$. For each conserved $k_\parallel$, the inverse
operator image $ \Asym(e^{i\theta}\U)_{k_\parallel}^{-1} \cdot
e^{ik_\perp x_\perp}$ can be similarly expanded and evaluated
explicitly as the sum of two complex sinusoids, leading to
\begin{widetext}
\begin{equation}
\rho_{sca} \leq \min_\theta \frac{1}{16\pi} \int_0^\infty \Re{ -\frac{
\tilde{\omega}^3 e^{i\theta} e^{2ik_\perp d}}{k_\perp^2} \left[ \frac{R_1}{r_1-ik_\perp} + \frac{R_2}{r_2-ik_\perp} \right]  +  \frac{\lvert\tilde{\omega}\rvert^3 e^{-2\Im(k_\perp)d}}{\lvert k_\perp \rvert^2} \left[\frac{R_1}{r_1+ik_\perp^*} + \frac{R_2}{r_2+ik_\perp^*} \right]  } \,\dd k_\parallel
\label{eq:halfspace_integral}
\end{equation}
\end{widetext}
where the closed-form complex coefficients
$R_{1,2}(k_\parallel;\theta)$ and decay constants
$r_{1,2}(k_\parallel;\theta)$ are given in the SI.

The main challenge in evaluating (\ref{eq:theta_dual}) comes from the
need to compute the inverse image of
$\Asym(e^{i\theta}\U)=\left[\Im{e^{i\theta}/\chi^*} +
  \Asym(e^{-i\theta}\G))\right]^{-1}$, which contains constraints on
wave propagation via its dependence on the vacuum Green's function
$\G$. In that regard, it is useful to consider an expansion of
(\ref{eq:theta_dual}) in orders of $\Asym(e^{-i\theta} \G)$ (after using the Cauchy-Schwartz inequality to relax the term involving $\vb{E}_{vac}^*$ (SI)):
\begin{align*}
    &\langle \rho \rangle_{max} = \frac{|\tilde{\omega}|}{2} \bra{\vb{E}_{vac}}\left[\Im{\frac{e^{i\theta}}{\chi^*}} + \Asym(e^{-i\theta}\G)\right]^{-1}\ket{\vb{E}_{vac}} \\
     &= \frac{|\tilde{\omega}|}{2} \bra{\vb{E}_{vac}} \Im{\frac{e^{i\theta}}{\chi^*}}^{-1} \left[\Id - \frac{\Asym(e^{-i\theta} \G)}{\Im{e^{i\theta}/\chi^*}} + \cdots \right] \ket{\vb{E}_{vac}} \\    
    &\leq \frac{|\tilde{\omega}|}{2} \Im{\frac{e^{i\theta}}{\chi^*}}^{-1} \braket{\vb{E}_{vac}} .\numthis
    \label{eq:Born}
\end{align*}
The zeroth-order approximation given in (\ref{eq:Born}) is especially
simple to evaluate as it only depends on $\chi$, and will hereafter be referred to as a material bound. A special yet sub-optimal value of $\theta$ in (\ref{eq:Born}) recovers prior LDOS bounds based on passivity (SI)~\cite{shim_fundamental_2019}.
Notably, this zeroth order approximation becomes loose whenever
$\Im{e^{i\theta}/\chi^*}^{-1} \Asym(e^{-i\theta} \G)$ is large as
compared to $\mathbb{I}$; this is the case for large $|\chi|$ or
large design domains (e.g., the half-space) where $\Asym\G$
dominates. Intuitively, terms containing the vacuum Green's function
capture physical limitations imposed by multiple scattering,
screening, and the finite speed of light, which limit achievable
LDOS.

Figure~\ref{fig:halfspace_lossless} shows upper bounds on the LDOS for
both TM and TE sources, obtained by evaluating
(\ref{eq:halfspace_integral}). In the near-field limit of $d \rightarrow
0$, the TM bounds are found to asymptote to a constant while the TE
bounds scale $\propto 1/d^2$, in agreement with a detailed asymptotic
analysis of the evanescent $k_\parallel \rightarrow \infty$ behavior
of the integrand in (\ref{eq:halfspace_integral}) (SI), and
matching prior results based on passivity~\cite{shim_fundamental_2019}. The constant
asymptote observed for TM sources as they approach the device reflects
a known artifact of 2d scalar electromagnetism, which precludes
non-integrable field singularities at sharp corners~\cite{andersen_field_1978}. In
contrast, the $1/d^2$ scaling of TE sources confirms the known
divergence in the energy concentration of a vector dipole in its near
field, with the exponent of $d$ tied to the number of spatial
dimensions; note that the maximum LDOS for a 3d source grows $\propto
1/d^3$ as $d \rightarrow 0$ (SI).

In the opposite far-field limit of $d \rightarrow \infty$ and for finite
$Q_{src}$, the bounds ultimately tend toward zero, reflecting a kind
of space--bandwidth constraint on the ability of structuring to affect
far-field emission over a finite bandwidth. One can, however, observe
plateaus ocurring at intermediate regimes, $\omega d /c \ll Q_{src}$,
wherein structuring can efficiently reflect traveling planewaves back
onto the dipole position. Achievable enhancements in this regime can
be shown to decay $\propto e^{-d/Q_{src}}$, manifesting as plateaus in
the various plots of Figure~\ref{fig:halfspace_lossless}b. Only in the
strictly single-frequency limit $Q_{src} \rightarrow \infty$ do
far-field LDOS contributions tend toward constants, $\rho_0/2$ and
$\rho_0$ for TE and TM sources, respectively, independently of
separation. The importance of properly capturing relevant wave
interference effects in these far-field regimes becomes evident when
analyzing corresponding predictions for material bounds~\cite{shim_fundamental_2019}, which
exhibit drastically inflated $\propto Q_{src}^2$ scaling in this
half-space setting (SI).

Asymptotic analysis of (\ref{eq:halfspace_integral}) supported by
Figure~\ref{fig:halfspace_lossless}b reveal that LDOS enhancements
saturate to finite values as $\abs{\chi} \rightarrow \infty$ (SI), in contrast to
prior bounds which grow indefinitely with increasing material
response~\cite{shim_fundamental_2019}. This is somewhat surprising given that larger $\chi$ implies a larger (potentially infinite) density of states within the material itself; ultimately, multiple scattering and screening effects lead to restrictions on the possible field localization at the source location that cannot be overcome with clever structuring. Similar conclusions have been reached concerning the efficacy of hyperbolic metamaterials~\cite{miller_effectiveness_2014} at enhancing the LDOS in that particular geometry class. 
It is also worth noting that the saturation characteristics of the
full bounds as a function of $d$ is distinct for the TM and TE
polarizations. For TM, reducing $d$ increases the relative advantage
of stronger materials; $\rho(\chi \rightarrow \infty) \propto
\log(1/d)$ as $d \rightarrow 0$ whereas $\rho(\chi<\infty)$ is finite as $d \rightarrow 0$ (SI). In contrast,
for TE the saturation behavior for large $|\chi|$ scales uniformly
with $d$; decreasing separation does not increase the relative
advantage of stronger materials.

\begin{figure}
\includegraphics[width=\linewidth]{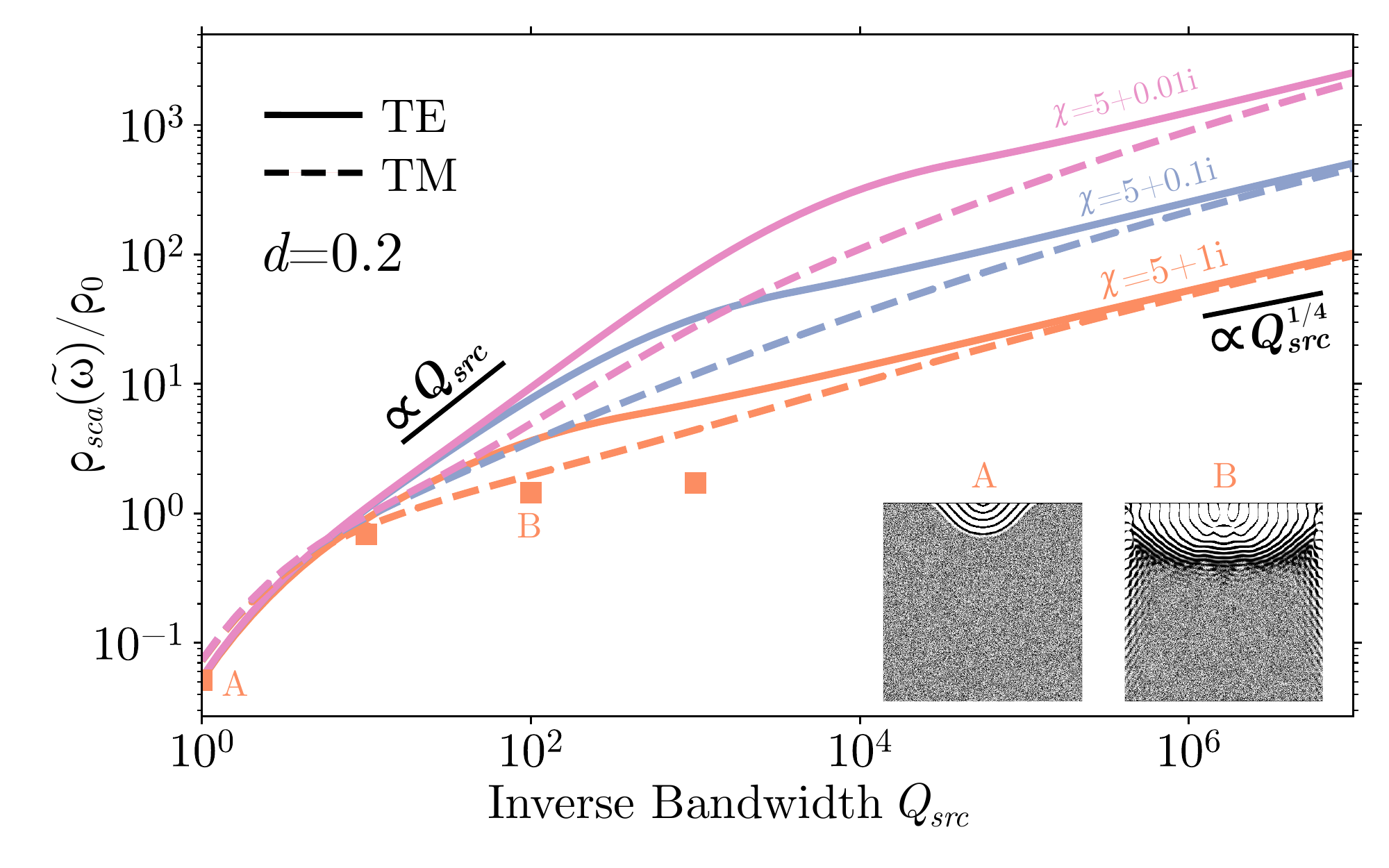}
\caption{\label{fig:halfspace_vary_Qs} Maximum LDOS enhancement near a
halfspace design region as a function of the source
  bandwidth. Squares are the performance of TM inverse designs over a
  $10 \times 10$ design region for $\chi=5+1i$, with insets $A$ and
  $B$ the structures found via TO for $Q_{src}=1$ and $Q_{src}=100$
  respectively.}
\end{figure}

With regard to bandwidth scaling, the broad structural freedom
afforded by an infinite design domain coupled with the lack of
material dissipation allows for creation of resonances with
arbitrarily small radiative loss rates, $Q_{mode} \to \infty$,
leading to the same $\propto Q_{src}$ dependence observed in
Figure~\ref{fig:FDFD_aggregate}. The situation changes in dissipative
media where mode lifetimes $Q_{mode} =
(\frac{1}{Q_{abs}}+\frac{1}{Q_{rad}})^{-1}$ contain both
radiative $Q_{rad}$ and absorptive $Q_{abs}$ contributions, with
$Q_{abs} \propto 1/\Im(\chi)$ denoting the proportion of
stored energy dissipated in the medium per unit cycle. Hence, under
finite absorption, $Q_{abs}$ sets a bound on the highest achievable
mode lifetime, with (\ref{eq:single_mode_avg}) suggesting a saturating
LDOS $\propto Q_{abs}$ as $Q_{src} \to \infty$.

In contrast, Figure~\ref{fig:halfspace_vary_Qs} shows that the bounds
exhibit a transition from the linear $\propto Q_{src}$ scaling of a
wide-bandwidth source coupled to a single resonance toward
\emph{diverging} response $\propto Q_{src}^{1/4}$ as $Q_{src} \to
\infty$, with the transition taking place as $Q_{src} \to Q_{abs}$. Intuitively, in
this regime wherein the lifetime of a single resonance is capped, one might expect the optimal design strategy to shift toward exploiting degeneracies. One class of geometry capable of achieving high modal degeneracies are waveguides, e.g., coupled cavities or photonic crystal gratings,
which at least in 1d are known to generate van Hove singularities in the
density of states~\cite{ibanescu_enhanced_2006}. Generally, a spectral singularity of the form
$\lvert \omega-\omega_0\rvert^{-\alpha}, \, 0<\alpha<1$ will yield a
Lorentzian-averaged LDOS that scales as
\begin{equation}
    \int_{-\infty}^\infty \frac{\Delta\omega_{src}/\pi}{(\omega-\omega_0)^2 +
      \Delta\omega_{src}^2} \frac{1}{\lvert \omega-\omega_0
      \rvert^\alpha}  \,\dd\omega \propto
    \frac{1}{(\Delta\omega_{src})^\alpha} \propto Q_{src}^\alpha .
\end{equation}
The bounds for the 2d half-space configuration thus suggest the
possibility of near-field absorbers supporting quartic ``band-edge''
dispersions capable of efficiently extracting evanescent fields,
resulting in singularities of the form $\lvert
\omega-\omega_0\rvert^{-1/4}$ (anomalous quartic "slow" group-velocity dispersions have been studied, for instance, in the context of waveguide solitons\cite{blanco-redondo_pure-quartic_2016}). Indeed, TO discovers structures
resembling adiabatic gratings (Figure~\ref{fig:halfspace_vary_Qs} insets) and exhibiting the
predicted quartic bandwidth scaling, though eventually saturating due to the
finite computational domain (SI).

\section{Summary}

We proposed an improved framework for evaluating upper bounds on the LDOS in structured media, and study in detail the effects of
source bandwidth, material susceptibility, and device footprint.  The
bounds provide a strong top-down complement to bottom-up design in at least two ways: First, they are useful in quantifying the optimality of existing "best" structures while diagnosing potential
areas of improvement: as seen in narrow bandwidth regimes, existing formulations of inverse
design fail to converge upon structures with performance close to the
global optimum, often outperformed by traditional ring resonators. Second, they allow studies of fundamental scaling characteristics without prior assumptions on device topology. 

For finite bandwidths, the quick saturation of the bounds with design 
size indicate that a device footprint of a few wavelengths is generally 
enough to achieve near-optimal performance. The observed saturation with increasing susceptibility has direct implications for material choice: a weak material
response can be mitigated given sufficient design size and
judicious structuring; conversely there are diminishing returns
associated with seeking large absolute values of the susceptibility,
raising the importance of other concerns such as material loss and
dispersion characteristics. The impact of material loss was also
evaluated in the context of maximum LDOS above a semi-infinite
structure, showing a transition from the intuitive linear $\propto
Q_{src}$ scaling expected of single-mode resonant enhancement to a
less obvious $\propto Q_{src}^{1/4}$ dependence given absorptive materials, inviting further studies into the
associated dispersion engineering mechanisms making such scaling
possible. While only results for dielectrics ($\Re(\chi)>0$) where shown in this paper, the framework can handle metals ($\Re(\chi)<0$) just as well, yielding broadly similar conclusions.

\section{Acknowledgments}
This work was supported by the National Science Foundation under the Emerging Frontiers in Research and Innovation (EFRI) program, Award No. EFMA-164098 and the Defense Advanced Research Projects
Agency (DARPA) under Agreements No. HR00111820046,
No. HR00112090011, and No. HR0011047197. The views,
opinions and findings expressed herein are those of the authors
and should not be interpreted as representing the official views
or policies of any institution. 
R.K.D. gratefully acknowledges financial support from the Princeton Presidential Postdoctoral Research Fellowship and from the National Academies of Science, Engineering, and Medicine Ford Foundation Postdoctoral Fellowship program.

\bibliography{LDOS}

\end{document}


\maketitle

\tableofcontents

\pagebreak

\section{Computational Details}
Inverse design in this work was performed using the ceviche Maxwell FDFD solver~\cite{hughes_forward-mode_2019} combined with the method of moving asymptotes algorithm as implemented in the non-linear optimization package NLOPT~\cite{johnson_nlopt_2019}. \\

Finite size LDOS bounds were computed using an in-house FDFD code to represent the Green's function and the electric fields over the design domain. The dual function is optimized with an in-house implementation of BFGS; for details on the computation of the dual gradient see~\cite{molesky_hierarchical_2020}.

\section{Dependence of bounds on number of subregion constraints}
\begin{figure}[!ht]
\includegraphics[width=\linewidth]{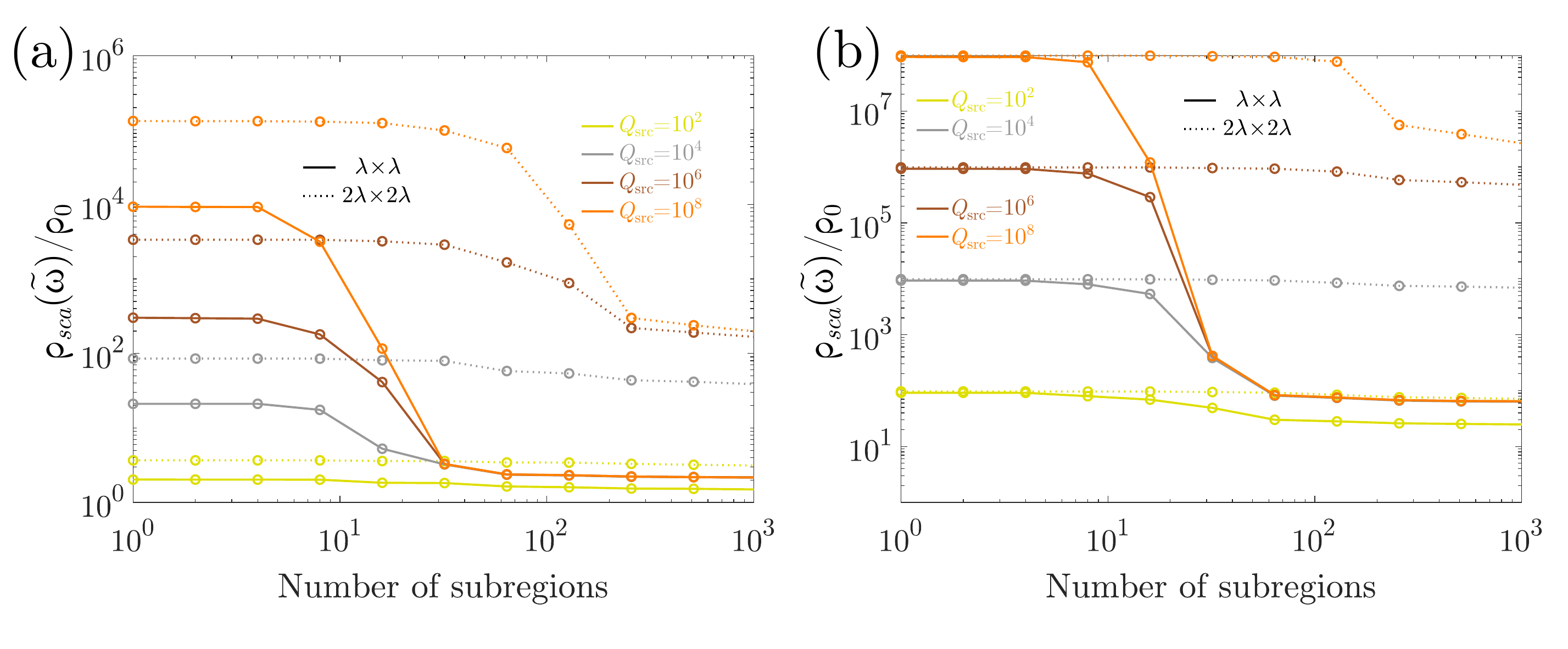}
\caption{\label{fig:subregion_convergence}  Plot showing the computed LDOS dual bounds as a function of the number of subregion constraints used. The subplots (a) and (b) correspond to the exterior dipole and interior dipole geometries shown in Fig. 1 of the main text. Solid lines are for system size $L=1$ and dotted lines are for system size $L=2$.}
\end{figure}

Fig. \ref{fig:subregion_convergence} shows how the bounds depend on the number of subregion constraints used. Generally there is a rapid transition of many orders of magnitude when the number of subregions reaches a critical value before convergence of the bounds. For larger $Q_{src}$ the transition is steeper, indicating the importance of subregion constraints for capturing the effects of radiative loss. For larger system size $L$, the transition is shallower: as stated in the main text, in the limit of large $L$ global constraints alone are sufficient to achieve tight bounds.

\pagebreak

\section{Ring resonator LDOS enhancement}
\begin{figure}[!ht]
\includegraphics[width=\linewidth]{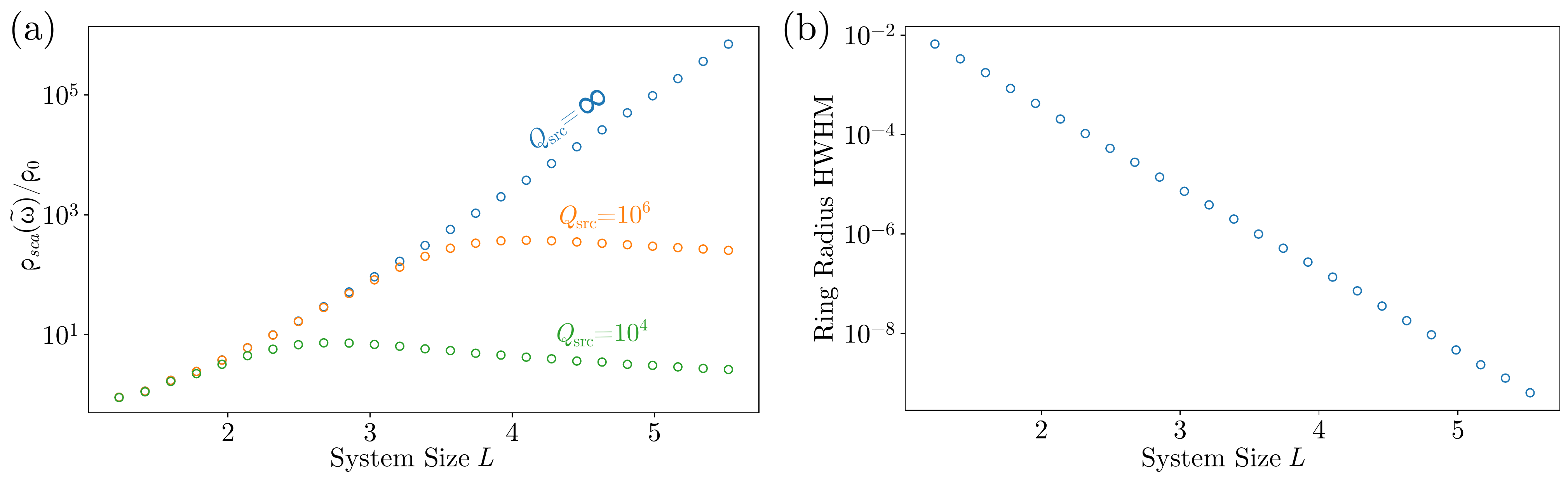}
\caption{\label{fig:ring resonator}  (a) LDOS enhancement of ring resonators with a fixed ring width of $0.2$ and TM dipole source separated $0.2$ away from the ring. The system size $L$ is 2 times the outer radius of the rings. (b) At $Q_{src}\rightarrow \infty$, the change in the ring radius $\Delta r$ such that the LDOS enhancement drops by a half. }
\end{figure}

In the main text we presented a specific ring resonator design with a ring width of $0.2$ that outperformed topology optimization from multiple random initializations. Here are extra data on the performance of ring resonators as a function of size for various $Q_{src}$. In the single frequency limit, we see the exponential scaling of LDOS enhancement as a function of ring size; for finite $Q_{src}$ there is a point where larger rings have higher $Q_{mode}$ but do not lead to larger bandwidth-averaged LDOS. For the $Q_{src}=10^6$ design shown in the main text the outer ring diameter is approximately $4.1$; Fig. \ref{fig:ring resonator}b indicates that we require around 8 significant figures of accuracy on the ring size to fall within $50\%$ of the maximum performance. 

\section{Bandwidth saturation of finite size LDOS bounds}
\begin{figure}[!ht]
\centering
\includegraphics[width=0.5\linewidth]{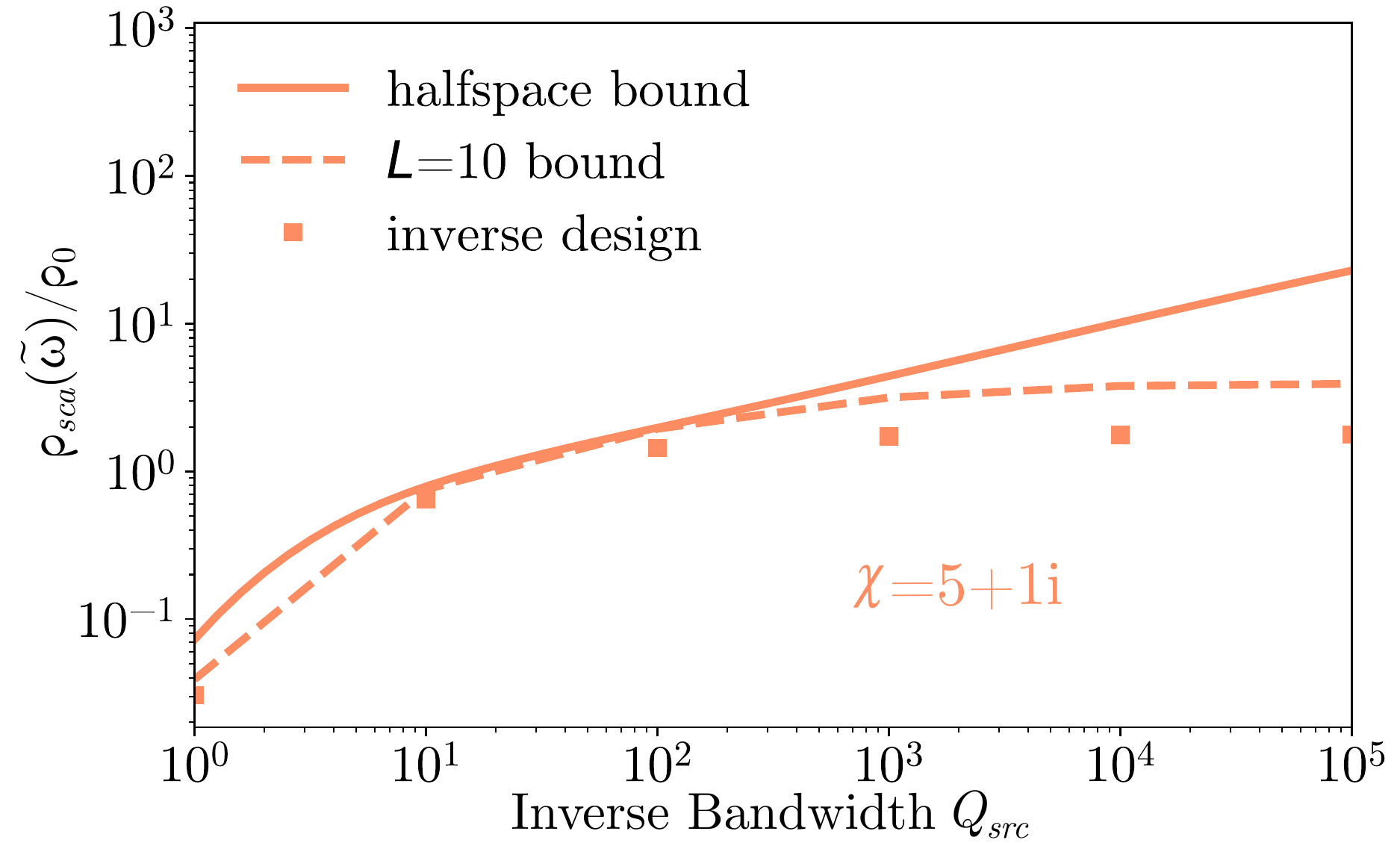}
\caption{\label{fig:finite_size}  LDOS bounds as a function of $Q_{src}$ for $\chi=5+1i$. }
\end{figure}

In Fig. 3 of the main text, the inverse designs are performed over a finite design domain of size $10$ by $10$, and the LDOS enhancement saturates with increasing $Q_{src}$. Fig. \ref{fig:finite_size} shows that this saturation is also present we you compute the bounds for the finite design domain; continued $Q_{src}^{1/4}$ scaling is seen just for the semi-infinite halfspace bounds.

\section{Differences in notation in supplementary information compared to the main text}
In the following sections, some quantities and variables use a different notation than what is written in the main text. This is predominantly for conciseness in writing down long derivations. For convenience a complete list of the differences in notation is listed here; each individual notation deviation will also be noted at its first occurrence in the SI.
\begin{itemize}
    \item The vacuum field $\vb{E}_{vac}$ in main text is referred to as $\vb{E}_v$ in the SI. 
    \item Use of $k_x$ and $k_y$ instead of $k_\parallel$ and $k_\perp$ for half-space bounds, with the $\hat{x}$ direction set as parallel to the half-space surface and $\hat{y}$ direction perpendicular to it. 
    \item the complex parameters $R_{1,2}$ and $r_{1,2}$ in the main text are replaced with $R_\pm$ and $r_\pm$ in the SI, where the $\pm$ subscripts have more background context within the derivation.
    \item With dimensionless units $c=1$, $\tilde{k} = \tilde{\omega}/c$ will often be used in place  $\tilde{\omega}$ in the context of spatial Fourier integrals. 
\end{itemize}

\section{Derivation of power conservation constraints via complex Poynting's theorem}

Here we present a derivation of the complex scattering constraints used via the complex Poynting's Theorem. For alternative derivations see early work [...].

\subsection{Complex Poynting's theorem for time harmonic fields and complex $\omega$}
In prior literature the complex Poynting's Theorem for time harmonic fields is presented given a real angular frequency $\omega$. The generalization to a complex $\omega$ is straightforward and written explicitly here for clarity. All fields have (complex) harmonic time dependence $e^{-i\omega t}$, with the harmonic time Maxwell's equations
\begin{subequations}
\begin{align}
    \div{\vb{D}} &= \rho_f \\
    \div{\vb{B}} &= 0 \\
    \curl{\vb{E}} &= i\omega \vb{B} \\
    \curl{\vb{H}} &= \vb{J}_f - i\omega\vb{D}
\end{align}
\end{subequations}
and linear constitutive relations $\vb{D}(\vb{r}) = \epsilon_0 \Teps_r(\vb{r}) \vb{E}(\vb{r})$ and $\vb{B}(\vb{r}) = \mu_0 \Tmu_r(\vb{r}) \vb{H}(\vb{r})$, with $\Teps(\vb{r})$ and $\Tmu(\vb{r})$ being tensor fields to allow for anisotropy. 
\\
Define the complex Poynting vector
\begin{equation}
    \vb{S} = \vb{E} \cross \vb{H}^*
\end{equation}
Taking its divergence yields
\begin{equation}
    \div{\vb{E}\cross\vb{H}^*} = \vb{H}^* \cdot \curl{\vb{E}} - \vb{E}\cdot\curl{\vb{H}^*}
\end{equation}
Now applying Maxwell's equations gives the differential form of the complex Poynting's theorem
\begin{align*}
    \div{\vb{E}\cross\vb{H}^*} &= i\omega \vb{H}^*\cross\vb{B} - i\omega^* \vb{E}\cdot\vb{D}^* - \vb{E}\cdot\vb{J}^* \\
    &= i\omega \vb{H}^* \cdot \Tmu \cdot \vb{H} - i\omega^* \vb{E}\cdot\Teps^*\cdot\vb{E}^* - \vb{E} \cdot \vb{J}^* \numthis
\end{align*}
The corresponding integral form is
\begin{equation}
\int_{\partial V} \dd\vb{\sigma} \cdot (\vb{E}\cross\vb{H}^*) = i\omega \int_V \vb{H}^*\cdot\Tmu\cdot\vb{H} \,\dd V - i\omega^* \int_V \vb{E}\cdot\Teps^*\cdot\vb{E}^* \,\dd V - \int_V \vb{E}\cdot\vb{J}^* \,\dd V
\end{equation}
in the case where $\omega$ is real we have
\begin{equation}
    \int_{\partial V} \dd\vb{\sigma} \cdot (\vb{E}\cross\vb{H}^*) = i\omega \int_V (\vb{H}^*\cdot\Tmu\cdot\vb{H} -  \vb{E}\cdot\Teps^*\cdot\vb{E}^*) \,\dd V - \int_V \vb{E}\cdot\vb{J}^* \,\dd V
\end{equation}
which is a more familiar form.

\subsection{Generalized energy conservation constraints}

To arrive at the generalized energy conservation constraints, we start with a scattering theory framework in which an initial free current source $\vb{J}_v$ produces the fields $\vb{E}_v$, $\vb{H}_v$ in vacuum. These initial fields interact with a scatterer, producing polarization currents $\vb{J}_s$ within the scatterer that generate scattered fields $\vb{E}_s$, $\vb{H}_s$. For simplicity, we assume that the scatterer is non-magnetic, i.e., $\Tmu = \mu_0$, and that the electric permittivity is local but may be anisotropic: $\Teps_r=\epsilon_0(\Id + \Id_s\Tchi\Id_s)$. The net result is a total field $\vb{E}_t = \vb{E}_v + \vb{E}_s$, $\vb{H}_t = \vb{H}_v + \vb{H}_s$. The complex Poynting theorem then applies in three settings: to the free current and initial fields $(\vb{J}_v, \vb{E}_v, \vb{H}_v)$ in vacuum, to the polarization current and scattered fields $(\vb{J}_s, \vb{E}_s, \vb{H}_s)$ in vacuum, and to the free current and total fields $(\vb{J}_v, \vb{E}_t, \vb{H}_t)$. \\

We consider as our region of interest some subset $V_k$ of the design region. For $(\vb{J}_v, \vb{E}_v, \vb{H}_v)$ in vacuum we have
\begin{equation}
    \int_{\partial V_k} \,\dd \vb{\sigma}\cdot(\vb{E}_v\cross\vb{H}_v) = i\omega \mu_0 \int_{V_k} \vb{H}_v^* \cdot \vb{H}_v \,\dd V - i\omega^* \epsilon_0 \int_{V_k} \vb{E}_v \cdot \vb{E}_v^* \,\dd V .
    \label{eq_Poynting_inc}
\end{equation}
For $(\vb{J}_s, \vb{E}_s, \vb{H}_s)$ in vacuum we have
\begin{equation}
    \int_{\partial V_k} \,\dd \vb{\sigma}\cdot(\vb{E}_s\cross\vb{H}_s) = i\omega \mu_0 \int_{V_k} \vb{H}_s^* \cdot \vb{H}_s \,\dd V - i\omega^* \epsilon_0 \int_{V_k} \vb{E}_s \cdot \vb{E}_s^* \,\dd V - \int_{V_k} \vb{E}_s \cdot \vb{J_s}^* \,\dd V .
    \label{eq_Poynting_sca}
\end{equation}

For $(\vb{J}_v, \vb{E}_t, \vb{H}_t)$ with the scatterer we have (noting that the free current is situated outside of the design region)
\begin{align*}
    &\int_{\partial V_k} \,\dd \vb{\sigma}\cdot((\vb{E}_v+\vb{E}_s)\cross(\vb{H}_v+\vb{H}_s)) \\
    &= i\omega \mu_0 \int_{V_k} (\vb{H}_v^* + \vb{H}_s^*) \cdot (\vb{H}_v + \vb{H}_s) \,\dd V - i\omega^* \epsilon_0 \int_{V_k} (\vb{E}_v+\vb{E}_s) \cdot (\Id + \Id_s\Tchi^*\Id_s) \cdot (\vb{E}_v^* + \vb{E}_s^*) \,\dd V . \numthis
    \label{eq_Poynting_tot}
\end{align*}
Subtracting (\ref{eq_Poynting_inc}) and (\ref{eq_Poynting_sca}) from (\ref{eq_Poynting_tot}) gives
\begin{align*}
    &\int_{\partial V_k} \,\dd \vb{\sigma}\cdot(\vb{E}_v\cross\vb{H}_s + \vb{E}_s\cross\vb{H}_v) \\
    = \, &i\omega \mu_0 \int_{V_k} (\vb{H}_v^* \cdot \vb{H}_s + \vb{H}_s^* \cdot \vb{H}_v) \,\dd V \\
    &- i\omega^*\epsilon_0 \int_{V_k} \big[ \vb{E}_v \cdot \Id_s\Tchi^*\Id_s \cdot \vb{E}_v^* + \vb{E}_s \cdot \Id_s\Tchi^*\Id_s \cdot \vb{E}_s^* + \vb{E}_v \cdot (\Id + \Id_s\Tchi^*\Id_s) \cdot \vb{E}_s^* + \vb{E}_s \cdot (\Id + \Id_s\Tchi^*\Id_s) \cdot \vb{E}_v^* \,\dd V \\
    &+ \int_{V_k} \vb{E}_s \cdot \vb{J_s}^* \,\dd V . \numthis
\end{align*}

To further simplify this expression, we investigate the cross term
\begin{equation*}
    \vb{H}_v^* \cdot \vb{H}_s = \frac{1}{\mu_0^2 |\omega|^2} (\curl{\vb{E}_s}) \cdot (\curl{\vb{E}_v^*}) = \frac{1}{\mu_0^2 |\omega|^2} \div(\vb{E}_s \cross \curl{\vb{E}_v^*}) + \frac{1}{\mu_0^2 |\omega|^2} \vb{E}_s \cdot \curl{\curl{\vb{E}_v^*}}
\end{equation*}
where we have used the identity
\begin{equation*}
    (\curl{\vb{A}})\cdot\vb{B} = \div(\vb{A}\cross\vb{B}) + \vb{A} \cdot (\curl{\vb{B}})  .
\end{equation*}
Now $\vb{E}_v$ satisfies the vector wave equation
\begin{align*}
    \curl{\curl{\vb{E}_v}} - \frac{\omega^2}{c^2} \vb{E}_v &= 0 \\
    \Rightarrow \curl{\curl{\vb{E}_v^*}} = \frac{\omega^{*2}}{c^2} \vb{E}_v^*
\end{align*}
over a source-free vacuum region, leading to
\begin{equation*}
    \vb{H}_v^* \cdot \vb{H}_s = \frac{1}{\mu_0^2 |\omega|^2} \div(\vb{E}_s \cross \curl{\vb{E}_v^*}) + \frac{\omega^*}{\mu_0^2 c^2 \omega} \vb{E}_s \cdot \vb{E}_v^* ,
\end{equation*}
\begin{align*}
    i \omega \mu_0 \int_{V_k} \vb{H}_v^* \cdot \vb{H}_s \,\dd V &= \frac{i}{\mu\omega^*} \int_{\partial V_k} \dd\sigma \cdot (\vb{E}_s \cross \curl{\vb{E}_v^*}) + i\omega^*\epsilon_0 \int_{V_k} \vb{E}_s \cdot \vb{E}_v^* \,\dd V \\
    &= \int_{\partial V_k} \dd\sigma \cdot (\vb{E}_s \cross \vb{H}_v^*) + i\omega^*\epsilon_0 \int_{V_k} \vb{E}_s \cdot \vb{E}_v^* \,\dd V .
\end{align*}
Thus we have the useful relation
\begin{equation}
    i \omega \mu_0 \int_{V_k} \vb{H}_v^* \cdot \vb{H}_s \,\dd V - \int_{\partial V_k} \dd\sigma \cdot (\vb{E}_s \cross \vb{H}_v^*) - i\omega^*\epsilon_0 \int_{V_k} \vb{E}_s \cdot (\Id + \Id_s\Tchi^*\Id_s) \cdot \vb{E}_v^* \,\dd V = -i\omega^*\epsilon_0 \int_{V_k} \vb{E}_s \cdot \Id_{s}\Tchi^*\Id_{s} \cdot \vb{E}_v^* \,\dd V .
\end{equation}
Similarly the other cross terms satisfy the relation
\begin{align*}
    &i \omega \mu_0 \int_{V_k} \vb{H}_v^* \cdot \vb{H}_s \,\dd V - \int_{\partial V_k} \dd\sigma \cdot (\vb{E}_s \cross \vb{H}_v^*) - i\omega^*\epsilon_0 \int_{V_k} \vb{E}_s \cdot (\Id + \Id_s\Tchi^*\Id_s) \cdot \vb{E}_v^* \,\dd V \\
    & = -i\omega^*\epsilon_0 \int_{V_k} \vb{E}_s \cdot \Id_s\Tchi^*\Id_s \cdot \vb{E}_v^* \,\dd V + \int_{V_k} \vb{E}_v \cdot \vb{J}_s^* \,\dd V \numthis
\end{align*}
where the additional $\int_{V_k} \vb{E}_v \cdot \vb{J}_s^* \,\dd V$ comes from from $\vb{E}_s$ satisfying a vector wave equation with the polarization currents as source:
\begin{equation*}
    \curl{\curl{\vb{E}_s}} - \frac{\omega^2}{c^2} \vb{E}_s = i\omega \mu_0 \vb{J}_s .
\end{equation*}
Substituting these cross-term relations into the expression and taking the complex conjugate gives

\begin{equation}
    \int_{V_k} \vb{E}_v^* \cdot \vb{J}_s \,\dd V = -i\omega \epsilon_0 \int_{V_k} \vb{E}_t^* \cdot \Id_s \Tchi \Id_s \cdot \vb{E}_t \,\dd V - \int_{V_k} \vb{E}_s^* \cdot \vb{J}_s \,\dd V .
\end{equation}
At this point, we can replace the polarization current $\vb{J}_s$, scattered electric field $E_s$, and total electric field $E_t$ with the polarization density $\vb{P}$ via the following relations
\begin{equation}
    \vb{J}_s = -i\omega \vb{P} \qquad \vb{P} = \epsilon_0 \Tchi \Id_s \cdot \vb{E}_t \qquad \vb{E}_s = \frac{1}{\epsilon_0} \G \cdot \vb{P}
\end{equation}
to finally obtain
\begin{equation}
    \int_{V_k} \vb{E}_v \cdot \bigg(\frac{\vb{P}}{\epsilon_0}\bigg) \,\dd V = \int_{V_k} \bigg(\frac{\vb{P}}{\epsilon_0}\bigg)^* \cdot  \bigg(\frac{1}{\chi^*} - \G^*\bigg) \cdot \bigg(\frac{\vb{P}}{\epsilon_0}\bigg) \,\dd V .
\end{equation}

\section{Lagrangian duality given global constraints}
In this section we detail how the dual problem for the two global power conservation constraints is equivalent to the dual problem for a single constraint with an additional parameter in the form of a complex phase rotation. 

We are interested in placing dual bounds on the power extracted from a dipole source given global conservation of power:
\begin{subequations}
\begin{align}
\text{maximize} \quad &\rho_{sca} = -\frac{1}{2} \Im{\tilde{\omega} \bra{\vb{E}^*_i} \ket{\vb{P}}} \\
\text{such that} \quad &\Re\bra{\vb{E}_v}\ket{\vb{P}} - \expval{\Sym\U}{\vb{P}} = 0 \\
&\Im\bra{\vb{E}_v}\ket{\vb{P}} - \expval{\Asym\U}{\vb{P}} = 0 .
\end{align}
\end{subequations}
The Lagrangian for this constrained optimization problem is
\begin{align*}
    \mathcal{L}(\vb{P}, \alpha_{Re}, \alpha_{Im}) &= -\frac{1}{2} \Im{\tilde{\omega}\bra{\vb{E}^*_i} \ket{\vb{P}}} + \alpha_{Re} \big[ \Re\bra{\vb{E}_v}\ket{\vb{P}} - \expval{\Sym\U}{\vb{P}} \big] + \alpha_{Im} \big[\Im\bra{\vb{E}_v}\ket{\vb{P}} - \expval{\Asym\U}{\vb{P}} \big] \\
    &= -\frac{1}{2} \Im{\tilde{\omega} \bra{\vb{E}^*_i} \ket{\vb{P}}} + \big(\frac{\alpha_{Re}}{2} + \frac{\alpha_{Im}}{2i}\big) \bra{\vb{E}_v}\ket{\vb{P}} + \big(\frac{\alpha_{Re}}{2} - \frac{\alpha_{Im}}{2i}\big) \bra{\vb{P}}\ket{\vb{E}_v} \\
    &+ \expval{\big[\big(\frac{\alpha_{Re}}{2} + \frac{\alpha_{Im}}{2i}\big)\U + \big(\frac{\alpha_{Re}}{2} - \frac{\alpha_{Im}}{2i}\big)\U^\dagger  \big]  }{\vb{P}} ,
\end{align*}
With the corresponding dual function
\begin{equation}
    \mathcal{D}(\alpha_{Re}, \alpha_{Im}) = \max_{\vb{P}} \mathcal{L}(\vb{P}, \alpha_{Re}, \alpha_{Im}) .
\end{equation}

Now define $\alpha \equiv \sqrt{\alpha_{Re}^2+\alpha_{Im}^2}$ and a complex phase rotation $p \equiv e^{i\theta} = (\alpha_{Im} + i\alpha_{Re})/\alpha$ the Lagrangian can be re-written as
\begin{align*}
    \mathcal{L}(\vb{P}, \alpha; \theta) &= -\frac{1}{2} \Im{\tilde{\omega} \bra{\vb{E}^*_i} \ket{\vb{P}}} + \alpha \bigg( \frac{e^{i\theta}\bra{\vb{E}^*_i}\ket{\vb{P}} - e^{-i\theta} \bra{\vb{P}}\ket{\vb{E}^*_i}}{2i} - \expval{ \frac{e^{i\theta}\U - e^{-i\theta}\U^\dagger}{2i} }{\vb{P}} \bigg) \\
    &= -\frac{1}{2} \Im{\tilde{\omega} \bra{\vb{E}^*_i} \ket{\vb{P}}} + \alpha \big[ \Im(p \bra{\vb{E}^*_i}\ket{\vb{P}}) - \expval{\Asym(p \U)}{\vb{P}} \big] \numthis
\end{align*}
which is exactly the Lagrangian of the single constraint optimization
\begin{subequations}
\begin{align}
\text{maximize} \quad & \rho_{sca} = -\frac{1}{2}\Im{\tilde{\omega}\bra{\vb{E}^*_i} \ket{\vb{P}}} \label{eq:theta_primal_objective}\\
\text{such that} \quad &\Im(p \bra{\vb{E}_v}\ket{\vb{P}}) - \expval{\Asym(p \U)}{\vb{P}} = 0 \label{eq:theta_primal_constraint}
\end{align}
\label{eq:theta_primal}
\end{subequations}
with corresponding dual function
\begin{equation}
    \mathcal{D}(\alpha; \theta) = \max_{\vb{P}} \mathcal{L}(\vb{P}, \alpha; \theta) .
    \label{def:theta_dual}
\end{equation}

Now it is clear that $(e^{i\theta}, \alpha)$ is just an alternate parametrization of the multiplier space of $(\alpha_{Re}, \alpha_{Im})$, hence the tightest dual bound is
\begin{equation}
    \min_{\alpha_{Re}, \alpha_{Im}} \mathcal{D}(\alpha_{Re}, \alpha_{Im}) = \min_\theta \min_\alpha \mathcal{D}(\alpha; \theta) .
\end{equation}

We can now derive an expression for $\min_\alpha \mathcal{D}(\alpha; \theta)$ for fixed phase rotation $\theta$. First, note that the Lagrangian $\mathcal{L}(\vb{P}, \alpha; \theta)$ only has a finite maximum when $\Asym(p\U) \succ 0$. See the following section for a spectral analysis of $\Asym(p\U)$ ascertaining the existence and range of complex phase rotation $p$ such that $\Asym(p\U) \succ 0$. We evaluate (\ref{def:theta_dual}) by calculating the stationary point of $\mathcal{L}$ with respect to $\ket{\vb{P}}$:
\begin{equation}
    \pdv{\mathcal{L}}{\bra{\vb{P}}} = 0 \quad \Rightarrow \quad \ket{\vb{P}} = -\frac{i \tilde{\omega}^*}{4\alpha} \Asym(p\U)^{-1} \ket{\vb{E}_v^*} + \frac{i}{2} p^* \Asym(p\U)^{-1} \ket{\vb{E}_v}. 
    \label{eq:theta_Popt}
\end{equation}
This stationary point maximizes $\mathcal{L}$ when $\Asym(p\U) \succ 0$. Given this positive definite condition, the primal problem (\ref{eq:theta_primal}) is also a convex problem with a non-empty feasible set, so strong duality holds [cite]. Thus to solve for the optimal $\bar{\alpha}$ that minimizes $\mathcal{D}(\alpha;\theta)$, we substitute (\ref{eq:theta_Popt}) into (\ref{eq:theta_primal_constraint}) and obtain
\begin{equation}
    \bar{\alpha} = \sqrt{\frac{\lvert \tilde{\omega} \rvert^2}{4\lvert p \rvert^2} \frac{\expval{\Asym(p\U)^{-1}}{\vb{E}_v^*} }{\expval{\Asym(p\U)^{-1}}{\vb{E}_v}} }  .
    \label{eq:theta_alphaopt}
\end{equation}
Now (\ref{eq:theta_Popt}) and (\ref{eq:theta_alphaopt}) can be simultaneously substituted back into (\ref{def:theta_dual}) to get the bound
\begin{align*}
    \rho_{sca} \leq \min_{\alpha} \mathcal{D}(\alpha; \theta) =& \, \frac{\lvert \tilde\omega \rvert}{4}  \sqrt{ \expval{\Asym(p\U)^{-1}}{\vb{E}_v^*} \expval{\Asym(p\U)^{-1}}{\vb{E}_v} } \\
    &- \frac{1}{4} \Re\Big\{ p^* \tilde{\omega} \bra{\vb{E}_v^*}\Asym(p\U)^{-1}\ket{\vb{E}_v} \Big\} \numthis
    \label{eq:theta_bound}
\end{align*}
as seen in the main text.

\section{Spectral analysis of Green's function}
Notation: in this and future sections of the supplementary info we will sometimes use the spatial wavevector $\tilde{k} = \tilde{\omega} / c = \tilde{\omega}$; in the main text only $\tilde{\omega}$ is used to reduce the number of different variables shown and to take advantage of the dimensionless units $c=1$. 

Starting from the planewave expansion of $\G$ as given by~\cite{tsang_scattering_2004} (multiplied by $\tilde{k}^2$ following our convention)
\begin{equation}
    \G = \frac{1}{(2\pi)^3} \iiint \dd^3\vb{k} e^{i\vb{k}\cdot(\vb{r}-\vb{r'})} \frac{\mathbb{I} \tilde{k}^2 - \vb{k}\otimes\vb{k}}{k^2-\tilde{k}^2},
\end{equation}
we have
\begin{equation}
    \Asym\G = \frac{1}{(2\pi)^3} \iiint \dd^3\vb{k} e^{i\vb{k}\cdot(\vb{r}-\vb{r'})} \bigg\{ \Im\bigg(\frac{\tilde{k}^2}{k^2-\tilde{k}^2}\bigg)\mathbb{I} - \Im\bigg(\frac{k^2}{k^2-\tilde{k}^2}\bigg) \unitv{k}\otimes\unitv{k} \bigg\} .
\end{equation}
Now $\tilde{k}^2 = \tilde{k}_r^2 - \tilde{k}_i^2 + 2i\tilde{k}_r\tilde{k}_i$; for notational convenience define $A_r\equiv\Re{\tilde{k}^2}=\tilde{k}_r^2-\tilde{k}_i^2$ and $A_i=\Im{\tilde{k}^2}=2\tilde{k}_r\tilde{k}_i$. Taking the imaginary part, we get
\begin{equation}
    \Asym\G = \frac{1}{(2\pi)^3} \iiint \dd^3\vb{k} e^{i\vb{k}\cdot(\vb{r}-\vb{r'})} \bigg\{ \frac{A_i k^2}{(k^2-A_r)^2 + A_i^2} \mathbb{I} - \frac{A_i k^2}{(k^2-A_r)^2 + A_i^2} \unitv{k}\otimes\unitv{k} \bigg\} .
\end{equation}
It is apparent now that the eigenwaves of $\Asym\G$ have the form $\unitv{e} e^{i\vb{k}\cdot\vb{r}}$, where $\unitv{e}$ is an eigenvector of the $3\times3$ matrix in parentheses. Under a triad that includes $\unitv{k}$, the $3\times3$ matrix is diagonal, and we see the longitudinal eigenwaves $\unitv{k} e^{i\vb{k}\cdot\vb{r}}$ have eigenvalue $0$. The transverse eigenwaves have eigenvalues
\begin{equation}
    \rho_t(k) = \frac{A_i k^2}{(k^2-A_r)^2+A_i^2} = \frac{2 \tilde{k}_r \tilde{k}_i k^2}{(k^2 - \tilde{k}_r^2 + \tilde{k}_i^2)^2 + (2 \tilde{k}_r \tilde{k}_i)^2} \geq 0
    \label{eq_AsymGeig} .
\end{equation}
Some observations about the scaling of these eigenvalues that are relevant for later discussion:
\begin{itemize}
    \item  $\Asym\G$ is positive semi-definite. The null-space always includes the longitudinal waves (which are irrelevant to the TM case). For single frequencies $\tilde{k}_i=0$ the null-space includes all evanescent waves; for finite bandwidth $\tilde{k}_i>0$ the null-space only contains extremely fast oscillations / extreme evanescent waves in the limit $|k|\rightarrow\infty$. 
    \item Some idea of the scaling of $\Asym\G^{-1}$ can be seen through the inverse of the transverse eigenvalues $1/\rho_t = \frac{1}{A_i} k^2 + (\frac{A_r^2}{A_i}+A_i)\frac{1}{k^2}-2\frac{A_r}{A_i}$, which for $k\geq \tilde{k}_r$ has a minimum around $k=|\tilde{k}|$ before scaling $\sim k^2$ as $k\rightarrow\infty$. 
\end{itemize}

\section{Spectral Analysis of $\Asym(p\U)$}
We have $\Asym(p\U) = \Asym(\chi^{-1\dagger}p) + \Asym(p^* \G)$. $\Asym(\chi^{-1\dagger}p)$ contributes a constant to all eigenvalues; we focus our attention on the second term $\Asym(p^* \G)$. Similar to the analysis in the previous subsection, we have
\begin{equation}
    \Asym\G = \frac{1}{(2\pi)^3} \iiint \dd^3\vb{k} e^{i\vb{k}\cdot(\vb{r}-\vb{r'})} \bigg\{ \Im\bigg(\frac{p^* \tilde{k}^2}{k^2-\tilde{k}^2}\bigg)\mathbb{I} - \Im\bigg(\frac{p^* k^2}{k^2-\tilde{k}^2}\bigg) \unitv{k}\otimes\unitv{k} \bigg\} .
\end{equation}
The longitudinal and transverse eigenvalues of $\Asym(p^* \G)$ come out to be
\begin{align*}
    \rho_{\G,l}(p) &= p_i ,\\
    \rho_{\G,t}(p) &= \frac{p_r k^2 A_i + p_i [-A_r(k^2-A_r) + A_i^2]}{(k^2-Ar)^2 + A_i^2} .
\end{align*}
The eigenvalues of $\Asym(p\U)$ are
\begin{align}
    \rho_{\U,l} &= \frac{\chi_i}{\chi_r^2+\chi_i^2} p_r + \bigg(1 + \frac{\chi_r}{\chi_r^2+\chi_i^2}\bigg) p_i ,\\
    \rho_{\U,t} &= \frac{p_r\chi_i + p_i\chi_r}{\chi_r^2 + \chi_i^2} + \frac{p_r k^2 A_i + p_i [-A_r(k^2-A_r) + A_i^2]}{(k^2-Ar)^2 + A_i^2} .
\end{align}
In order for the dual bound (\ref{eq:theta_bound}) to be valid we need $\Asym(p\U)$ to be positive definite, i.e., $\rho_{\U}>0$. 

\subsection{Lossless dielectrics}
For lossless dielectrics we have $\chi_i=0$, $\chi_r>0$. The PD condition $\rho_{\U,l}>0$ then leads to $p_i>0$. Setting $p_r=1$, we require
\begin{equation}
    \rho_{\U,t} = \frac{1}{\chi_r} p_i + \frac{k^2 A_i + p_i [-A_r(k^2-A_r) + A_i^2]}{(k^2-A_r)^2 + A_i^2} > 0.
\end{equation}
Defining for convenience $u \equiv k^2$, the problem is then finding
\begin{equation}
    \min_{u \geq 0} f(u) = \frac{(A_i - p_i A_r)u + p_i(A_r^2+A_i^2)}{(u-A_r)^2 + A_i^2} .
\end{equation}

It is clear that $f(0) = p_i>0$ and $\lim_{u\rightarrow\infty} f(u) = 0$. In the special case $p_i = A_i/A_r$, the only critical point is $u^* = A_r$ which is a maximum, so $\min f(u) = 0$ and $\rho_{\U,t} = \frac{1}{\chi_r} p_i > 0$. When $p_i \neq A_i/A_r$ the critical points are (assuming $A_i \ll A_r$)
\begin{equation}
    u^{*\pm} \approx \frac{p_i A_r^2 \pm \sqrt{1+p_i^2} A_i A_r}{p_i A_r - A_i} .
\end{equation}
If $p_i < A_i/A_r$ then $u^{*+}<0$ and irrelevant, while $u^{*-}$ is a maximum, so $\min f(u)=0$. If $p_i>A_i/A_r$, then $u^{*+}$ is the minimum, with
\begin{equation}
    \min f(u) = f(u^{*+}) \approx \frac{-\sqrt{1+p_i^2} A_r A_i}{\bigg[ \frac{(1+\sqrt{1+p_i^2}) A_r A_i}{p_i A_r - A_i} \bigg]^2 + A_i^2} < 0 .
\end{equation}
Thus for lossless dielectrics, $\Asym(p\U)$ is PD given $p_r=1$ and $0<p_i\leq A_i / A_r$. For specific values of $\chi_r$, larger rotation angles with $p_i>A_i / A_r$ is possible. By how much? Set
\begin{equation}
    p_i = c \cdot \frac{A_i}{A_r} ;
\end{equation}
now, assume that $p_i \ll 1$ in the limit $A_i \rightarrow 0$ for vanishing bandwidth, this assumption can be checked for consistency at the end. This yields
\begin{equation}
    \min f(u) \approx - \frac{(c-1)^2}{4} \frac{A_i}{A_r} .
\end{equation}
Then
\begin{equation}
    \rho_{\U,t}>0 \Rightarrow \frac{p_i}{\chi_r} > \frac{(c-1)^2}{4} \frac{A_i}{A_r} \Rightarrow c^2 - \bigg(\frac{4}{\chi_r} + 2\bigg)c + 1 < 0 .
\end{equation}
Conclude that
\begin{equation}
    c < 1 + 2\bigg[\frac{1}{\chi_r} + \sqrt{\frac{1}{\chi_r^2} + \frac{1}{\chi_r}} \bigg] ,
    \label{eq:cmax}
\end{equation}
so $c_{max}$ is a monotonic function of $\chi_r$ and not dependent on the bandwidth, $p_{i,max} \propto A_i \ll 1$ in the limit $A_i \rightarrow 0$. 

\section{Global constraint bounds for 2D TM dipole near a half-space}
In this section we evaluate (\ref{eq:theta_bound}) explicitly for the case of 2D TM dipole source $\vb{J}(x',y') = \delta(x')\delta(y'+d)$ a distance $d$ away from a half-space design region $V=\{(x,y)|y>0\}$. 

Notation: in the main text the $\parallel$ and $\perp$ symbols were used to indicate directions parallel and perpendicular to the surface of the half-space design region, respectively. In this section we use explicit Cartesian coordinates with $x$ being $\parallel$ and $y$ being $\perp$. 

\subsection{TM Green's function and dipole field}

Following \cite{tsang_scattering_2004}, the 2D TM Green's function is
\begin{equation}
    \G^{TM}(\vb{\rho},\vb{\rho'}) = \tilde{k}^2 \frac{i}{4\pi}
    \begin{cases}
    \int_{-\infty}^\infty \dd k_x \frac{\hat{z}\hat{z}}{k_y} e^{i k_x(x-x')}e^{ik_y(y-y')} & y>y' \\
    \\
    \int_{-\infty}^\infty \dd k_x \frac{\hat{z}\hat{z}}{k_y} e^{i k_x(x-x')}e^{-ik_y(y-y')} & y<y'
    \end{cases}
    \label{def:TM_Green}
\end{equation}
where $\tilde{k} = \tilde{k}_r + i\tilde{k}_i$ is complex for finite bandwidths, $k_y=\sqrt{\tilde{k}^2-kx^2}$ always taking the root with positive imaginary part.

The dipole source $\vb{J}(x',y') = \delta(x')\delta(y'+d)$ produces a vacuum electric field within the $y>0$ design region
\begin{subequations}
\begin{equation}
    \vb{E}_v^{TM}(x,y) = \int_{-\infty}^\infty \frac{e^{i k_x x}}{\sqrt{2\pi}} \vb{E}_{v,k_x}^{TM}(y) \,\dd k_x, 
\end{equation}
\begin{equation}
    \vb{E}_{v,k_x}^{TM}(y) = -\frac{\tilde{k}}{2\sqrt{2\pi}} \frac{1}{k_y} e^{i k_y y} e^{i k_y d} .
    \label{eq:TM_Ei}
\end{equation}
\end{subequations}
The single frequency vacuum dipole radiation is
\begin{equation}
    \rho_0^{TM} = \frac{\Re{\tilde{k}}}{8} = \frac{\pi}{4 \lambda_0} .
\end{equation}
The composite $\Asym(p\U^{TM})$ integral operator is
\begin{subequations}
\begin{equation}
    \Asym(p\U^{TM})(x,y,x',y') = \int_{-\infty}^\infty \dd k_x \frac{e^{i k_x x}}{\sqrt{2\pi}} \Asym(p\U^{TM})_{k_x}(y,y') \frac{e^{-i k_x x'}}{\sqrt{2\pi}}
\end{equation}
with
\begin{equation}
    \Asym(p\U^{TM})_{k_x} (y,y') = \Im{p/\chi^*} \delta(y-y') + \frac{1}{2} \Re{p^* \frac{\tilde{k}^2}{k_y} e^{i k_y |y-y'|}} .
    \label{eq:TM_AsympUkx}
\end{equation}
\end{subequations}

\subsection{Simplification of (\ref{eq:theta_bound})}
For the half-space design region, (\ref{eq:theta_bound}) may be simplified by observing that $\expval{\Asym(p\U^{TM})^{-1}}{\vb{E}_v^{TM}} = \expval{\Asym(p\U^{TM})^{-1}}{\vb{E}_v^{TM*}}$. We have
\begin{equation}
    \expval{\Asym(p\U^{TM})^{-1}}{\vb{E}_v^{TM}} = \frac{\abs{\tilde{k}}^2}{8\pi} \int_{-\infty}^\infty \dd k_x  \frac{e^{-2k_{yi} d}}{\abs{k_y}^2} \iint_0^\infty e^{-ik_y^* y} \Asym(p\U^{TM})_{k_x}^{-1}(y,y') e^{i k_y y} \, \dd y' \dd y
    \label{eq:Ei_AsympUinv_Ei}
\end{equation}
where we have made multiple use of the Fourier completeness relation $\frac{1}{2\pi} \int_{-\infty}^\infty e^{i(k_x - k_x') x} \,\dd x = \delta(k_x - k_x')$. 
Now,
\begin{align*}
    \vb{E}_v^{TM*}(x,y) &= -\frac{\tilde{k}^*}{2\sqrt{2\pi}} \int_{-\infty}^\infty  \frac{e^{-ik_x x}}{\sqrt{2\pi}} \frac{1}{k_y^*} e^{-i k_y^* y} e^{-i k_y^* d} \, \dd k_x \\
    &= -\frac{\tilde{k}^*}{2\sqrt{2\pi}} \int_{-\infty}^\infty  \frac{e^{ik_x x}}{\sqrt{2\pi}} \frac{1}{k_y^*} e^{-i k_y^* y} e^{-i k_y^* d} \, \dd k_x \qquad kx \rightarrow -kx \numthis
\end{align*}
where in the second line we have made use of the fact that $k_y = \sqrt{\tilde{k}^2 - k_x^2}$ is invariant to the sign of $k_x$. This gives
\begin{equation}
    \expval{\Asym(p\U^{TM})^{-1}}{\vb{E}_v^{TM}} = \frac{\abs{\tilde{k}}^2}{8\pi} \int_{-\infty}^\infty \dd k_x \frac{e^{-2k_{yi} d}}{\abs{k_y}^2} \iint_0^\infty e^{i k_y y} \Asym(p\U^{TM})_{k_x}^{-1}(y,y') e^{-i k_y^* y} \, \dd y' \dd y.
    \label{eq:Ei*_AsympUinv_Ei*}
\end{equation}
Now from (\ref{eq:TM_AsympUkx}) we see that $\Asym(p\U^{TM})_{k_x}(y,y')$ is invariant under exchange of the $y$, $y'$ dummy integration variables. Since (\ref{eq:Ei_AsympUinv_Ei}) and (\ref{eq:Ei*_AsympUinv_Ei*}) are related by the exchange of $y$, $y'$, conclude that 
\begin{equation}
\expval{\Asym(p\U^{TM})^{-1}}{\vb{E}_v^{TM}} = \expval{\Asym(p\U^{TM})^{-1}}{\vb{E}_v^{TM*}}. 
\label{eq:Ei*_Ei_equality}
\end{equation}
This allows us to simplify (\ref{eq:theta_bound}) to
\begin{equation}
    \rho_{sca}^{TM} \leq \frac{\abs{\tilde{\omega}}}{4} \expval{\Asym(p\U^{TM})^{-1}}{\vb{E}_v^{TM}} - \frac{1}{4} \Re\Big\{ p^* \tilde{\omega} \bra{\vb{E}_v^{TM*}} \Asym(p\U)^{-1} \ket{\vb{E}_v^{TM}} \Big\} .
    \label{eq:theta_bound_simplified}
\end{equation}
For certain asymptotic and approximation analyses, we will also make use of (\ref{eq:Ei*_Ei_equality}) and the Cauchy-Schwartz inequality to relax the second term in (\ref{eq:theta_bound_simplified}), yielding
\begin{equation}
    \rho_{sca}^{TM} \leq \frac{\abs{\tilde{\omega}}}{2} \expval{\Asym(p\U^{TM})^{-1}}{\vb{E}_v^{TM}}
    \label{eq:theta_bound_simplified_CS}
\end{equation}

\subsection{Evaluation of $\Asym(p\U^{TM})^{-1} \ket{\vb{E}_v^{TM}}$}
The core calculation in (\ref{eq:theta_bound_simplified}) is the evaluation of 
\begin{equation}
    \Asym(p\U^{TM})^{-1} \ket{\vb{E}_v^{TM}} = -\frac{\tilde{k}}{2\sqrt{2\pi}} \int_{-\infty}^\infty \dd k_x \frac{e^{i k_x x}}{\sqrt{2\pi}} \frac{e^{i k_y d}}{k_y} \int_0^\infty \Asym(p\U^{TM})_{k_x}^{-1}(y,y') e^{i k_y y'} \, \dd y'.
    \label{eq:AsympUinv_Evac}
\end{equation}
Defining
\begin{equation}
    f(y) = \int_0^\infty \Asym(p\U^{TM})_{k_x}^{-1}(y,y') e^{i k_y y'} \, \dd y',
\end{equation}
we have
\begin{equation}
    \int_0^\infty \Asym(p\U^{TM})_{k_x}(y,y') f(y') \,\dd y' = e^{i k_y y'}.
    \label{eq:TM_Laplace_target}
\end{equation}

The action of $\Asym(p\U)_{k_x}$ is a convolution over $y$; this suggests that the Laplace transform $\mathscr{L}\{h(t)\}(s) = \int_0^\infty h(t) e^{-st} \,\dd t$ may be used to solve (\ref{eq:TM_Laplace_target}). We now take the Laplace transform of both sides of (\ref{eq:TM_Laplace_target}), defining $F(s) \equiv \mathscr{L}\{f(y)\}$ and the auxiliary variable $B \equiv p^* \frac{\tilde{k}^2}{k_y}$. The LHS gives
\begin{align*}
\Im{\frac{p}{\chi^*}} F(s) &+ \frac{B_r+iB_i}{4} \mathscr{L}\bigg\{ \int_0^y e^{-(k_{yi}-ik_{yr})(y-y')} f(y')\,\dd y'  \bigg\} + \frac{B_r-iB_i}{4} \mathscr{L}\bigg\{ \int_0^y e^{-(k_{yi}+ik_{yr})(y-y')} f(y') \,\dd u' \bigg\} \\
&+ \frac{B_r+iB_i}{4} \mathscr{L} \bigg\{ \int_y^\infty e^{(k_{yi}-ik_{yr})(y-y')} f(y') \,\dd y' \bigg\} + \frac{B_r-iB_i}{4} \mathscr{L} \bigg\{ \int_y^\infty e^{(k_{yi}+k_{yr})(y-y')} f(y') \,\dd y' \bigg\}.
\end{align*}
The RHS is simply
\begin{equation*}
    \mathscr{L}\bigg\{ e^{(-k_{yi}+ik_{yr})y} \bigg\} = \frac{1}{s+(k_{yi}-ik_{yr})} .
\end{equation*}
Evaluating the LHS, the Laplace transform of a convolution gives
\begin{equation}
    \Laplace{\int_0^y e^{-a (y-y')} f(y')\,\dd y' } = \frac{F(s)}{s+a} .
    \label{eq:Laplace_conv}
\end{equation}
The transform of the other type of integral can be evaluated explicitly to a simple form:
\begin{align*}
    \Laplace{\int_y^\infty e^{a (y-y')} f(y') \,\dd y'} &= \int_0^\infty e^{-sy} \int_y^\infty e^{a(y-y')} f(y') \,\dd y' \,\dd y\\
    &= \int_0^\infty e^{-ay'} f(y') \bigg( \int_0^{y'} e^{-(s-a)y} \, \dd y \bigg) \, \dd y' \qquad \text{exchange order of integration} \\
    &= \int_0^\infty f(y') e^{-ay'} \frac{1}{a-s} \bigg( e^{(a-s)y'} -1 \bigg) \, \dd y' \\
    &= \frac{1}{a-s} \bigg( \int_0^\infty f(y') e^{-sy'} \,\dd y' - \int_0^\infty f(y') e^{-ay'} \,\dd y' \bigg) \\
    &= \frac{1}{a-s} ( F(s) - F(a) ) . \numthis
    \label{eq:Laplace_cross_corr}
\end{align*}
Thus $F(s)$ satisfies
\begin{align*}
    \frac{1}{s+(k_{yi}-ik_{yr})} = \frac{p_i}{\chi} Fs) &+ \frac{Br+iB_i}{4} \frac{F(s)}{s+(k_{yi}-ik_{yr})} \\
    &+ \frac{B_r - iB_i}{4} \frac{F(s)}{s+(k_{yi}+ik_{yr})} \\
    &+ \frac{B_r+iB_i}{4} \frac{1}{(k_{yi}-ik_{yr})-s} \bigg[ F(s) - F(k_{yi}-ik_{yr})\bigg] \\
    &+ \frac{B_r - iB_i}{4} \frac{1}{(k_{yi}+ik_{yr})-s} \bigg[ F(s) - F(k_{yi}+ik_{yr}) \bigg] . \numthis
\end{align*}

Solving this will give us an explicit form of $F(s)$ dependent on two free parameters $F(k_{yi}-ik_{yr})$ and $F(k_{yi}+ik_{yr})$. This seemingly gives, instead of a single solution $f(y)$, a whole family of solutions. However, as we shall see, only one member of this family belongs to $L^2[0,\infty)$, decaying as $y \rightarrow \infty$. The other members of this family grow exponentially as $y \rightarrow \infty$ which violates the requirements of Fubini's theorem \cite{rudin_real_2006} needed to justify the order of integration exchange used in (\ref{eq:Laplace_cross_corr}), and are not true solutions to (\ref{eq:TM_Laplace_target}). 

Solving for $F(s)$ yields 
\begin{subequations}
\begin{equation}
    F(s) = \frac{F_{num}(s)}{F_{denom}(s)}
\end{equation}
where the numerator is
\begin{align*}
    F_{num}(s) = [s-(k_{yi}-ik_{yr})][s^2-(k_{yi}+ik_{yr})^2] &- \frac{Br+iB_i}{4} \gamma_- [s+(k_{yi}-ik_{yr})][s^2-(k_{yi}+ik_{yr})^2] \\
    &- \frac{B_r-iB_i}{4} \gamma_+ [s+(k_{yi}+ik_{yr})][s^2-(k_{yi}-ik_{yr})^2] \numthis
\end{align*}
and the denominator is
\begin{equation}
    F_{denom}(s) = \Im{\frac{p}{\chi^*}} s^4 - \bigg[ 2\Im{\frac{p}{\chi^*}} (k_{yi}^2-k_{yr}^2) + (B_r k_{yi} + B_i k_{yr})\bigg]s^2 + \Im{\frac{p}{\chi^*}} \abs{k_y}^4 + (B_r k_{yi} - B_i k_{yr}) \abs{k_y}^2 ,
\end{equation}
\end{subequations}
where for notational convenience, we have defined $\gamma_{\pm} \equiv F(k_{yi} \pm ik_{yr})$. 

The roots $r$ of $F_{denom}(s)$ are then the poles of $F(s)$:
\begin{subequations}
\begin{align}
    &(+,-)r_+ = (+,-)\bigg\{(k_{yi}^2 - k_{yr}^2) + \frac{1}{2\Im{p/\chi^*}} \bigg[(B_r k_{yi} + B_i k_{yr}) + \sqrt{\Delta} \,\bigg] \bigg\}^{1/2} ,\\
    &(+,-)r_- = (+,-)\bigg\{(k_{yi}^2 - k_{yr}^2) + \frac{1}{2\Im{p/\chi^*}} \bigg[(B_r k_{yi} + B_i k_{yr}) - \sqrt{\Delta} \,\bigg] \bigg\}^{1/2} ,\\
    &\Delta=(B_r k_{yi} + B_i k_{yr})^2 + 8\Im{\frac{p}{\chi^*}} \Bigg( B_i k_{yr} k_{yi}^2 - B_r k_{yr}^2 k_{yi} - \Im{\frac{p}{\chi^*}} k_{yr}^2 k_{yi}^2 \Bigg),
\end{align}
\begin{equation}
    F_{denom}(s) = \Im{\frac{p}{\chi^*}} (s-r_+)(s+r_+)(s-r_-)(s+r_-) .
\end{equation}
\label{eq:TM_halfspace_poles}
\end{subequations}

We see that the poles come in two pairs $\pm r_+$ and $\pm r_-$. In general, one each from $(+,-) r_+$ and $(+,-) r_-$ will have a positive real part and its opposite sign counterpart will have a negative real part; we take $r_+$ and $r_-$ to be the poles with positive real parts (in the main text these are $r_1$ and $r_2$ respectively. The inverse Laplace transform is
\begin{equation}
    f(y) = \mathscr{L}^{-1}\bigg\{ F(s) \bigg\} = \frac{1}{2\pi i} \int_{T-i\infty}^{T+i\infty} F(s) e^{sy} \,\dd s
\end{equation}
where $T \in \mathbb{R}$ is greater than the real parts of all the poles of $F(s)$. Since $y>0$, we can deform the line integration and complete the contour in the $\Re{s}<0$ half plane upon which the contour completion part decays to 0; thus the inverse Laplace transform picks out the residue of $F(s) e^{sy}$. It is clear then that the contribution to the residue from $r_+$ and $r_-$ leads to functions with exponential growth as $y \rightarrow \infty$ whereas the contribution from $-r_+$ and $-r_+$ lead to functions with exponential decay as $y \rightarrow \infty$. To recover $f(y) \in L^2[0,\infty)$ we need to select free parameters $\gamma_\pm$ such that the residue for $r_+$ and $r_-$ are 0. This leads to the following linear system of equations from which we solve for $\gamma_\pm$:
\begin{subequations}
\begin{align*}
    &[r_+ - (k_{yi}-ik_{yr})][r_+^2 - (k_{yi}+ik_{yr})^2] \\
    &= \frac{B_r+iB_i}{4}[r_+ + (k_{yi}-ik_{yr})][r_+^2 - (k_{yi}+ik_{yr})^2]\gamma_- + \frac{Br-iB_i}{4}[r_+ + (k_{yi}+ik_{yr})][r_+^2 - (k_{yi}-ik_{yr})^2]\gamma_+ , \numthis \\
    &[r_- - (k_{yi}-ik_{yr})][r_-^2 - (k_{yi}+ik_{yr})^2] \\
    &= \frac{B_r+iB_i}{4}[r_- + (k_{yi}-ik_{yr})][r_-^2 - (k_{yi}+ik_{yr})^2]\gamma_- + \frac{Br-iB_i}{4}[r_- + (k_{yi}+ik_{yr})][r_-^2 - (k_{yi}-ik_{yr})^2]\gamma_+ . \numthis \\
\end{align*}
\label{eq:TM_halfspace_gamma}
\end{subequations}
$f(y)$ then takes on the form
\begin{equation}
    f(y) = R_+ e^{-r_+ y} + R_- e^{-r_- y}
    \label{eq:AsympUinv_expiky}
\end{equation}
with the coefficients of the exponentials ($R_1$ and $R_2$ in the main text) given by
\begin{align*}
    R_\pm = \frac{1}{\mp 2\Im{p/\chi^*} r_\pm \cdot (r_+^2-r_-^2)}
    \bigg\{&[-r_\pm - (k_{yi}-ik_{yr})][r_\pm^2-(k_{yi}+ik_{yr})^2] \\
    &- \frac{B_r+iB_i}{4} [-r_\pm + (k_{yi}-ik_{yr})][r_\pm^2-(k_{yi}+ik_{yr})^2]\gamma_- \\
    &- \frac{B_r-iB_i}{4}[-r_\pm + (k_{yi}+ik_{yr})][r_\pm^2-(k_{yi}-ik_{yr})^2]\gamma_+
    \bigg\} .\numthis
    \label{eq:TM_halfspace_residues}
\end{align*}

(\ref{eq:AsympUinv_expiky}) can now be substituted back into (\ref{eq:AsympUinv_Evac}) and (\ref{eq:theta_bound_simplified}) to get an analytical integral expression for the LDOS enhancement bound near a half-space:

\begin{equation}
    \rho_{sca} \leq \frac{1}{16\pi} \int_0^\infty \Re{ -\frac{\tilde{k}^3 p \, e^{2ik_y d}}{k_y^2} \left( \frac{R_+}{r_+ - ik_y} + \frac{R_-}{r_- - ik_y} \right) + \frac{\abs{\tilde{k}}^3 e^{-2 k_{yi} d}}{\abs{k_y}^2} \left( \frac{R_+}{r_+ + ik_y^*} + \frac{R_-}{r_- + ik_y^*} \right) }
    \label{eq:TM_halfspace_kxintegral}
\end{equation}

\subsection{TE Green's function and dipole field}
The 2D TE Green's function is
\begin{equation}
    \G^{TE}(x,y,x',y') = -\hat{y}\hat{y}\delta(x-x')\delta(y-y') + \tilde{k}^2 \frac{i}{4\pi}
    \begin{cases}
    \int_{-\infty}^\infty \dd k_x \frac{\hat{h}(k_y)\hat{h}(k_y)}{k_y} e^{i k_x(x-x')}e^{ik_y(y-y')} & y>y' \\
    \\
    \int_{-\infty}^\infty \dd k_x \frac{\hat{h}(-k_y)\hat{h}(-k_y)}{k_y} e^{i k_x(x-x')}e^{-ik_y(y-y')} & y<y'
    \end{cases}
\end{equation}
with the unit vectors $\hat{h}$ given by
\begin{equation}
    \hat{h}(k_y) = -\frac{k_y}{\tilde{k}}\hat{x} + \frac{k_x}{\tilde{k}}\hat{y} \qquad \hat{h}(-k_y) = \frac{k_y}{\tilde{k}}\hat{x} + \frac{k_x}{\tilde{k}}\hat{y}. 
\end{equation}

A dipole source of the form $\vb{J}^y(x',y') = \hat{y} \delta(x')\delta(y'+d)$ produces a vacuum field
\begin{subequations}
\begin{equation}
    \vb{E}_v^{TEy}(x,y) = \int_{-\infty}^\infty \frac{e^{i k_x x}}{\sqrt{2\pi}} \vb{E}_{v,k_x}^{TEy}(y) \, \dd k_x, 
\end{equation}
\begin{equation}
    \vb{E}_{v,k_x}^{TEy}(y) = -\frac{\sqrt{2\pi}}{4\pi \tilde{k}} \cdot e^{ik_y d} e^{ik_y y} \cdot \left(-k_x \hat{x} + \frac{k_x^2}{k_y} \hat{y} \right) .
\end{equation}
\end{subequations}
The vacuum dipole radiation is
\begin{equation}
    \rho_0^{TEy} = \frac{\Re{\tilde{k}}}{16} = \frac{\pi}{8\lambda_0} .
\end{equation}
The composite $\Asym(p\U^{TE})$ integral operator is
\begin{multline}
    \Asym(p\U^{TE}) = \int_{-\infty}^\infty \dd k_x \frac{e^{ik_x(x-x')}}{2\pi} \Bigg\{ \Im{\frac{p}{\chi^*}} \delta(y-y')(\hat{x}\hat{x}+\hat{y}\hat{y}) +  \delta(y-y')p_i \hat{y}\hat{y} + \\
    \frac{1}{2} \Re{p^* k_y e^{ik_y\abs{y-y'}}} \hat{x}\hat{x} 
    + \frac{1}{2} \Re{p^*(k_x^2/k_y)e^{i k_y\abs{y-y'}}}\hat{y}\hat{y} - \frac{1}{2}\text{sgn}(y-y') \Im{p^* k_x e^{i k_y \abs{y-y'}}} (\hat{x}\hat{y} + \hat{y}\hat{x})  \Bigg\}. 
    \label{eq:TE_AsympU}
\end{multline}

The corresponding global constraint bound for a TE dipole source near a half-space can be derived following an analogous procedure to that of a TM dipole source, though the expressions involved are tedious and was done in practice using Mathematica. For conciseness, the expressions are not reproduced here. 

\section{Asymptotic analysis}
Based on the results from the prior Section, we can do asymptotic analysis to understand the observed scaling of the half-space bounds with regards to bandwidth, material, and separation. 

\subsection{TM bandwidth scaling}
To understand the scaling of the LDOS bounds with bandwidth, it is illuminating to consider the contributions from the traveling waves and evanescent waves separately. Here traveling waves and evanescent waves refer to different $k_x$ in the planewave decomposition of the dipole field: traveling waves have $|k_x|<\tilde{k}_r$ and a predominantly real $k_y$, propagating for long distances with decay rate proportional to the bandwidth $\tilde{k}_i$; evanescent waves have $|k_x|>\tilde{k}_r$ and a predominantly imaginary $k_y$, rapidly decaying along the $y$ direction with decay rate strongly dependent on $|k_x|$.

\subsubsection{Traveling wave contribution}
We show that in the limit of zero bandwidth / single frequency, the contribution from just the traveling waves to the LDOS limits tends toward a finite constant. The following analysis is for TM fields but an analogous result can be obtained for TE fields.\\
We take the material susceptibility $|\chi|\rightarrow\infty$ and phase rotation $p=1$ so the results are material-independent and $\Asym(p\U^{TM})=\Asym\G^{TM}$:
\begin{align*}
    \Asym\G^{TM}(x,y;x',y') &= \int_{-\infty}^\infty \frac{e^{ik_x(x-x')}}{2\pi} \Asym\G_{k_x}^{TM}(y,y') \,\dd k_x ,\\
    \Asym\G_{k_x}^{TM}(y,y') &= \frac{1}{2} \Re{\frac{\tilde{k}^2}{k_y} e^{ik_y|y-y'|}}  .
\end{align*}
In the limit of zero bandwidth / single frequency, $\tilde{k}\in\mathbb{R}$. Then for $|k_x|>\tilde{k}$, $k_y$ is purely imaginary and $\Asym\G_{k_x}^{TM}=0$, leading to
\begin{align*}
    \Asym\G^{TM} &= \int_{-\tilde{k}}^{\tilde{k}} \,\dd k_x \frac{e^{ik_x(x-x')}}{2\pi} \frac{\tilde{k}^2}{k_y} \cos k_y(y-y') \\
    &= \int_{-\tilde{k}}^{\tilde{k}} \,\dd k_x \frac{e^{ik_x(x-x')}}{2\pi} \frac{\tilde{k}^2}{k_y} \big[\cos(k_y y)\cos(k_y y') + \sin(k_y y)\sin(k_y y') \big] .
\end{align*}

We consider a design region shaped as a slab with infinite extent in the $x$ direction and a thickness $h$ in the $y$ direction with the origin at the center. This is different from a half-space, but as we shall see the final result is completely independent of $h$ and thus should apply in the limit $h \rightarrow \infty$. The different parities of $\cos$ and $\sin$ about the origin allow us to directly write down the eigenvectors of $\Asym\G_{k_x}$ as
\begin{subequations}
\begin{align}
    \ket{q_{cos} (k_y)} &= \frac{\cos(k_y y)}{a_{cos}} ,\\
    \ket{q_{sin} (k_y)} &= \frac{\sin(k_y y)}{a_{sin}}
\end{align}
with normalization factor
\begin{align}
    a^2_{cos} &= \int_{-h/2}^{h/2} \cos^2(k_y y) \,\dd y = \frac{1}{2k_y}(k_y h + \sin(k_y h)) ,\\
    a^2_{sin} &= \int_{-h/2}^{h/2} \sin^2(k_y y) \,\dd y = \frac{1}{2k_y}(k_y h - \sin(k_y h)) 
\end{align}
and corresponding eigenvalues
\begin{align}
    \rho_{cos} (k_y) &= \frac{\tilde{k}^2}{2k_y} a_{cos}^2 ,\\
    \rho_{sin} (k_y) &= \frac{\tilde{k}^2}{2k_y} a_{sin}^2 .
\end{align}
\end{subequations}
We can thus write
\begin{equation}
    \Asym\G^{TM} = \int_{-\tilde{k}}^{\tilde{k}} \,\dd k_x \frac{e^{ik_x(x-x')}}{2\pi} \bigg( \rho_{cos}(k_y) \ket{q_{cos}(k_y)}\bra{q_{cos}(k_y)} + \rho_{sin}(k_y) \ket{q_{sin}(k_y)}\bra{q_{sin}(k_y)} \bigg) .
\end{equation}

The dipole source $\vb{J}(x',y') = \delta(x')\delta(y'-d-h/2)$ generates an incident field with the traveling wave components given by
\begin{align*}
    \vb{E}_{v,travel}^{TM} &= -\frac{\tilde{k}}{2\sqrt{2\pi}} \int_{-\tilde{k}}^{\tilde{k}} \dd k_x \frac{e^{ik_x x}}{\sqrt{2\pi}} \frac{1}{k_y} e^{ik_y(d+h/2)} \big( \cos(k_y y) - i \sin(k_y y) \big) ,\\
    -\vb{E}_{v,travel}^{TM*} &= \frac{\tilde{k}}{2\sqrt{2\pi}} \int_{-\tilde{k}}^{\tilde{k}} \dd k_x \frac{e^{ik_x x}}{\sqrt{2\pi}} \frac{1}{k_y} e^{-ik_y(d+h/2)} \big( \cos(k_y y) + i \sin(k_y y) \big) .
\end{align*}
While $\Asym\G$ is technically semi-definite, the incident field $\vb{E}_{v,travel}$ is contained completely in the span of the non-singular eigenvectors $q_{cos}(k_y)$, $q_{sin}(k_y)$. Thus we can write down $\Asym\G^{-1} \cdot \vb{E}_{v,travel}$ as
\begin{equation}
    \Asym\G^{TM-1} \cdot \vb{E}_{v,travel}^{TM} = -\frac{\tilde{k}}{2\sqrt{2\pi}} \int_{-\tilde{k}}^{\tilde{k}} \dd k_x \frac{e^{ik_x x}}{\sqrt{2\pi}} \frac{1}{k_y} e^{ik_y(d+h/2)} \big( \rho_{cos}^{-1} \cos(k_y y) - i \rho_{sin}^{-1} \sin(k_y y) \big)
\end{equation}
which can be formally interpreted as a pseudo-inverse or the inverse restricted to the relevant non-singular subspace. Thus we have
\begin{align*}
    \bra{\vb{E}_{v,travel}^{TM}}\Asym\G^{TM-1}\ket{\vb{E}_{v,travel}^{TM}} &= \frac{ \tilde{k}^2 }{8\pi} \int_{-\tilde{k}}^{\tilde{k}} \frac{1}{k_y^2} \bigg(\frac{a_{cos}^2}{\rho_{cos}} + \frac{a_{sin}^2}{\rho_{sin}} \bigg) \,\dd k_x \\
    &= \frac{ \tilde{k}^2 }{8\pi} \int_{-\tilde{k}}^{\tilde{k}} \frac{4}{\tilde{k}^2 k_y} \,\dd k_x \\
    &= \frac{1}{2\pi} \int_{-\tilde{k}}^{\tilde{k}} \frac{1}{\sqrt{\tilde{k}^2-k_x^2}} \,\dd k_x \\
    &= \frac{1}{2} .\numthis
\end{align*}
A similar calculation yields
\begin{align*}
    \bra{-\vb{E}_{v,travel}^*}\Asym\G^{-1}\ket{\vb{E}_{v,travel}} &= \frac{ \tilde{k}^2}{8\pi} \int_{-\tilde{k}}^{\tilde{k}} \frac{e^{ik_y(2d+h)}}{k_y^2} \bigg(\frac{a_{cos}^2}{\rho_{cos}} - \frac{a_{sin}^2}{\rho_{sin}} \bigg) \,\dd k_x \\
    &= 0 .\numthis
\end{align*}
This gives the traveling wave contribution to the material-independent LDOS enhancement limits
\begin{align*}
    \rho_{sca}^{TM} &=  \frac{1}{4} \Re{ \bra{-\vb{E}_{v,travel}^{TM*}} (\Asym\G^{TM})^{-1}\ket{\vb{E}_{v,travel}^{TM}} } + \frac{\tilde{k}}{4} \bra{\vb{E}_{v,travel}^{TM}} (\Asym\G^{TM})^{-1}\ket{\vb{E}_{v,travel}^{TM}}  \\
    &= \frac{\tilde{k}}{8} .\numthis
\end{align*}
Thus, if only the traveling waves (far field) are considered, $\rho_{sca}^{TM}/\rho_0^{TM}$ tends towards a constant value of $1$ in the single frequency limit. This can indeed be observed in the numerics for narrow bandwidth and large separation $d$, where the near-field contribution has decayed exponentially to the point of being negligible. An analogous calculation for TE shows that there the ratio is $1/2$.

\subsubsection{Evanescent wave contribution\label{sec:TM_evan} }
We evaluate the bandwidth scaling of the evanescent wave contribution for lossless materials ($\chi_i=0$) and TM polarization. To do so, we choose the phase parameter $p = 1+ip_i$ with the imaginary part $p_i \rightarrow 0$. As we shall see, this not only simplifies the asymptotic analysis for small bandwidth but also will lead to material independent bounds that are finite. 

Given $\chi_i=0$, we have the material factor $\lim_{p_i\rightarrow 0} \Im{p/\chi^*} \rightarrow 0$. Furthermore, from (\ref{eq:TM_halfspace_poles}) we have
\begin{equation}
    r_- \sim \sqrt{k_{yi}^2 - k_{yr}^2 - \frac{2 k_{yr}k_{yi} (B_i k_{yi} - B_r k_{yr})}{B_r k_{yi} + B_i k_{yr}} } \qquad r_+ \sim \sqrt{\frac{\chi (B_r k_{yi} + B_i k_{yr})}{p_i}} \qquad p_i \rightarrow 0
\end{equation}
so we see that in the limit $p_i \rightarrow 0$, $r_-$ tends toward a finite value while $r_+$ diverges $\propto p_i^{-1/2}$. 

To simplify the analysis, we can use the Cauchy-Schwartz relaxed (\ref{eq:theta_bound_simplified_CS}) to get
\begin{align*}
    \rho_{sca}^{TM} &\leq \frac{\abs{\tilde{\omega}}}{2} \expval{\Asym(p\U^{TM})^{-1}}{\vb{E}_v^{TM}} \\
    &= \frac{1}{8\pi} \int_0^\infty \Re{ \frac{\abs{\tilde{k}}^3 e^{-2 k_{yi} d}}{\abs{k_y}^2} \left( \frac{R_+}{r_+ + ik_y^*} + \frac{R_-}{r_- + ik_y^*} \right) } \,\dd k_x . \numthis
    \label{eq:TM_halfspace_kxintegral_CS}
\end{align*}
Now, treating $r_+$ as a large variable in the limit $p_i \rightarrow 0$, asymptotic analysis of (\ref{eq:TM_halfspace_gamma}) and (\ref{eq:TM_halfspace_residues}) gives
\begin{equation}
    \frac{R_+}{r_+ + ik_y^*} + \frac{R_-}{r_- + ik_y^*} \sim \frac{8 k_{yi} \abs{k_y}^2}{\Im{p/\chi^*} r_+^2 \left[(r_-+k_{yi})^2 +k_{yr}^2 \right]} \qquad p_i \rightarrow 0
\end{equation}
Since $\Im{p/\chi^*} \propto p_i$ and $r_+^2 \propto p_i^{-1}$ in the limit $p_i$, we see that the entire integrand of (\ref{eq:TM_halfspace_kxintegral_CS}) tends toward a finite value
\begin{equation}
    \rho_{sca}^{TM} \leq \frac{1}{8\pi} \int_0^\infty \Re{ \abs{\tilde{k}}^3 e^{-2 k_{yi} d} \frac{8 k_{yi}}{(B_r k_{yi} + B_i k_{yr}) \left[(r_-+k_{yi})^2 +k_{yr}^2 \right]}  } \,\dd k_x ,
    \label{eq:TM_halfspace_kxintegrand_CS_mat_indep}
\end{equation}
so the bound with $p=1$ is well-defined. We can now consider the narrow bandwidth limit $\tilde{k}_i \rightarrow 0$ given $p=1$. It has already been established that the traveling wave contribution from the $k_x<\tilde{k}_r$ part of the integrand is a constant in the narrow bandwidth limit. In the evanescent region $kx > \tilde{k}_r$, to lowest order in $\tilde{k}_i$ we have
\begin{subequations}
\begin{equation}
    k_{yi} \sim \sqrt{k_x^2 - \tilde{k_r}^2} \quad k_{yr} \sim \frac{\tilde{k}_r}{k_{yi}} \cdot \tilde{k}_i \qquad \tilde{k}_i \rightarrow 0
    \label{eq:evan_ky}
\end{equation}
\begin{equation}
    B_i \sim -\frac{\tilde{k}_r^2}{k_{yi}} \quad B_r \sim \left( \frac{\tilde{k}_r^3}{k_{yi}^3} + \frac{2\tilde{k}_r}{k_{yi}} \right) \cdot \tilde{k}_i \qquad \tilde{k}_i \rightarrow 0
\end{equation}
\begin{equation}
    r_-(p=1) \sim \sqrt{\tilde{k}_r^2 + k_{yi}^2} \qquad \tilde{k}_i \rightarrow 0
\end{equation}
\label{eq:TM_halfspace_smallki}
\end{subequations}
All of these combine together to give
\begin{equation}
    \Re{ \abs{\tilde{k}}^3 e^{-2 k_{yi} d}  \frac{8 k_{yi}}{(B_r k_{yi} + B_i k_{yr}) \left[(r_-+k_{yi})^2 +k_{yr}^2 \right]}  } \propto \tilde{k}_i^{-1} \qquad kx>\tilde{k}_r, \tilde{k}_i \rightarrow 0
\end{equation}
and we see that the evanescent wave contribution gives the inverse bandwidth scaling for lossless materials, as seen in the main text. 

\subsection{TM material Scaling}
From section \ref{sec:TM_evan} we see that given lossless $\chi$ the bound with $p=1$ is finite and well-defined, amounting to
\begin{equation}
    \rho_{sca}^{TM} \leq \frac{\tilde{\omega}}{2} \expval{\Asym\G^{TM-1}}{\vb{E}_v^{TM}},
    \label{eq:TM_mat_indep}
\end{equation}
which is a bound independent of $\chi$, and thus therefore applies to all lossless $\chi$. The next step is to show that (\ref{eq:TM_mat_indep}) is also a bound for lossy $\chi$ with $\Im(\chi)>0$. We can see this by again setting $p=1$ in (\ref{eq:theta_bound_simplified}) and applying the Cauchy-Schwartz inequality to arrive at a bound for lossy $\chi$:
\begin{equation}
    \rho_{sca}^{TM} \leq \frac{\abs{\tilde{\omega}}}{2} \expval{\left(\frac{\chi_i}{\abs{\chi}^2} + \Asym\G^{TM}\right)^{-1}}{\vb{E}_v^{TM}} .
\end{equation}
Now as operators we have $\frac{\chi_i}{\abs{\chi}^2} \succ 0$ and $\Asym\G^{TM} \succeq 0$, so $\expval{\left(\frac{\chi_i}{\abs{\chi}^2} + \Asym\G^{TM}\right)^{-1}}{\vb{E}_v^{TM}} < \expval{\Asym\G^{TM-1}}{\vb{E}_v^{TM}}$ and it is clear that the material independent bound (\ref{eq:TM_mat_indep}) while derived assuming $\chi_i=0$ applies to general lossy $\chi$. Thus in general as $|\chi| \rightarrow \infty$, the TM halfspace LDOS limits do not diverge, but are always bounded by (\ref{eq:TM_mat_indep}). 

\subsection{TM separation scaling}
In this section we investigate the scaling of the TM half-space bounds with the vacuum separation $d$ as $d \rightarrow 0$. The $k_x$ integral in  (\ref{eq:TM_halfspace_kxintegral_CS}) converges for finite $d$ due to the exponential decay factor $e^{-2 k_{yi} d}$: as $k_x \rightarrow \infty$, $k_{yi} \approx k_x \rightarrow \infty$. The scaling of (\ref{eq:TM_halfspace_kxintegral_CS}) with $d$ thus depends on the $k_x$ scaling of the rest of the integrand as $k_x \rightarrow \infty$. 

\subsubsection{Finite $\chi$ \label{sec:asymptotics_TM_finitechi}}
Given finite $\chi$, we can always select a complex phase $p$ such that $\Im{p/\chi^*} > 0$. Now from (\ref{eq:TM_AsympUkx}) and (\ref{eq:evan_ky}) it is clear that $\lim_{k_x \rightarrow \infty} \Asym(p\U^{TM})_{k_x} = \Im{p/\chi^*}$, since $\lim_{k_x \rightarrow \infty} \tilde{k}^2/k_y = 0$ and $\lim_{k_x \rightarrow \infty} e^{ik_y|y-y'|} \rightarrow 0$ for $y \neq y'$ (if $y=y'$ then $e^{ik_y|y-y'|}=1$ but this is a finite non-zero value of the integral kernel with support over a set of measure 0 and the integral operator as a whole still goes to 0). 

The $k_x$ integrand of $\expval{\Asym(p\U^{TM})^{-1}}{\vb{E}_v^{TM}}$ given in (\ref{eq:Ei_AsympUinv_Ei}) for large $k_x$ is then approximately
\begin{equation*}
    \frac{e^{-2k_x d}}{k_x^2} \Im{p/\chi^*}^{-1} \int_0^\infty e^{-2k_x y} \, \dd y \propto \frac{e^{-2k_x d}}{k_x^3}.
\end{equation*}
Even at exactly $d=0$, the integrand scales $\propto k_x^{-3}$ as $k_x \rightarrow \infty$ and the integral converges. Thus for finite $\chi$ the half-space LDOS bounds approach a finite constant as $d \rightarrow 0$. 

\subsubsection{Material independent bounds}
For the material independent bound given in (\ref{eq:TM_mat_indep}), the $k_x$ integrand is given by (\ref{eq:TM_halfspace_kxintegrand_CS_mat_indep}) and (\ref{eq:TM_halfspace_smallki}). In the narrow-bandwidth, large $k_x$ limit the integrand
\begin{equation}
    \Re{ \abs{\tilde{k}}^3 e^{-2 k_{yi} d}  \frac{8 k_{yi}}{(B_r k_{yi} + B_i k_{yr}) \left[(r_-+k_{yi})^2 +k_{yr}^2 \right]} } \sim \frac{\abs{\tilde{k}}^3}{k_r} \frac{1}{k_i} \frac{e^{-2k_x d}}{k_x}  \qquad k_i \rightarrow 0, k_x \rightarrow \infty \qquad k_x \rightarrow \infty.
\end{equation}
Thus the material independent bound scaling as $d \rightarrow 0$ can be evaluated as
\begin{align*}
    \rho_{sca, max}^{TM} \sim \frac{1}{8\pi} \frac{\abs{\tilde{k}}^3}{k_r} \frac{1}{k_i} \int_{\beta \tilde{k}_r}^{\infty} \frac{e^{-2k_x d}}{k_x} \,\dd k_x \\
    = \frac{1}{8\pi} \frac{\abs{\tilde{k}}^3}{k_r} \frac{1}{k_i} E_1(2\beta \tilde{k}_r d) \\
    \sim \frac{1}{8\pi} \frac{\abs{\tilde{k}}^3}{k_r} \frac{1}{k_i} \ln(\lambda_0 / d) \numthis
    \label{eq:TM_halfspace_d_asymptotics_mat_indep}
\end{align*}
where we have selected a constant $\beta \gg 1$ such that the large $k_x$ approximations hold; as seen in (\ref{eq:TM_halfspace_d_asymptotics_mat_indep}) the exact value of $\beta$ is irrelevant to the leading order asymptotic of $\ln(1/d)$. Note that this is a different scaling than the finite $\chi$ bounds in the previous subsection. A consequence of the material independent bounds diverging whilst fixed material bounds saturating as $d \rightarrow 0$ is that smaller $d$ increases the relative advantage of large materials, as seen in Fig. (2b) in the main text. 

\subsection{TE separation scaling}
The prior sections on the asymptotics of the TM case have demonstrated scaling  of the bounds with the inverse of the bandwidth (for lossless materials) and saturation with increasing material susceptibility; the TE bounds share these scaling characteristics so for the sake of simplicity we will not do a detailed TE asymptotic analysis. 

However, there is a difference in the $d \rightarrow 0$ scaling between the TM and TE results, which can be understood by comparing the $k_x \rightarrow \infty$ characteristics of $\vb{E}_{v,k_x}^{TM}$ and $\vb{E}_{v,k_x}^{TEy}$. For TE, we have
\begin{align*}
    \rho_{sca}^{TEy} &\leq 2 \int_0^\infty \expval{\Asym(p\U_{k_x}^{TE})^{-1}}{\vb{E}_{v,k_x}^{TEy}} \, \dd k_x\\
    &= 2 \int_0^\infty \expval{(\Im{p/\chi^*} + \Asym(p^*\G_{k_x}^{TE}))^{-1}}{\vb{E}_{v,k_x}^{TEy}} \, \dd k_x \\
    & < 2 \int_0^\infty \Im{p/\chi^*}^{-1} \lVert \vb{E}_{v,k_x}^{TEy} \rVert^2 \, \dd k_x \\
    &= \frac{1}{4\pi \abs{\tilde{k}}^2} \int_0^\infty \frac{e^{-2 k_{yi}d}}{2 k_{yi}} \left(k_x^2 + \frac{k_x^4}{\abs{k_y}^2} \right) \, \dd k_x .
\end{align*}
In the $k_x \rightarrow \infty$ limit, the integrand $\sim k_x e^{-2 k_x d}$, so similar to (\ref{sec:asymptotics_TM_finitechi}), for the $d \rightarrow 0$ asymptotics we have
\begin{align*}
    \rho_{sca}^{TEy} &\lesssim \frac{1}{4\pi \abs{\tilde{k}}^2} \int_{\beta \tilde{k}_r}^\infty k_x e^{-2 k_x d} \, \dd k_x \qquad d \rightarrow 0 \\
    &= \frac{1}{d^2} \frac{1}{4\pi \abs{\tilde{k}}^2} \cdot \frac{1}{4} e^{-2\beta \tilde{k}_r d} (2 \beta \tilde{k}_r d  + 1) \qquad d\rightarrow 0 \\
    &\propto \frac{1}{d^2}  \qquad d \rightarrow 0 \numthis \label{eq:TE_halfspace_d_asymptotics}
\end{align*}
and we recover the $1/d^2$ scaling of the TE bounds seen in the main text.

\subsubsection{3D separation scaling}
For a point dipole near a 3D half-space design region, a similar analysis would yield $1/d^3$ scaling of the LDOS bounds. Due to the extra spatial dimension as compared with the TE case, the 1D $k_x$ integral becomes a 2D $k_\parallel$ integral. For simplicity, suppose that the dipole is aligned with the normal of the design region surface, giving the problem cylindrical symmetry. The spatial integration kernel is then $2\pi \int_0^\infty k_\parallel (\cdot) \dd k_\parallel$, with the bound integral analogous to (\ref{eq:TE_halfspace_d_asymptotics}) having an extra $k_\parallel$ factor in the integrand. This will lead to $\propto 1/d^3$ scaling, as seen in \cite{shim_fundamental_2019}.

\section{Comparison to passivity bounds}
The main prior work concerning finite bandwidth LDOS limits is \cite{shim_fundamental_2019}, where an bound based on passivity constraints was derived:
\begin{subequations}
\begin{equation}
    \rho_{sca} \leq \frac{|\tilde{\omega}|}{2} f(\tilde{\omega}, \chi) \bra{\vb{E}_v}\ket{\vb{E}_v}
\end{equation}
with the material figure of merit
\begin{equation}
    f(\tilde{\omega}, \chi) = \frac{|\tilde{\omega}\chi|^2 + |\tilde{\omega}\chi|\Delta\tilde{\omega}}{|\tilde{\omega}|\Im(\tilde{\omega}(1+\chi))}.
    \label{eq:passivity_FOM}
\end{equation}
\label{eq:passivity_bounds}
\end{subequations}
Note that compared to (22) in \cite{shim_fundamental_2019}, we have dropped the electrostatic contribution (as in the main text), and there is an additional factor of $\pi |\tilde{\omega}|^2$ due to differences in definitions (the $\pi$ factor is a matter of convention; the $|\tilde{\omega}|^2$ factor comes from $\vb{E}_v$ in our work coming from a unit amplitude dipolar current source instead of a unit dipole moment). 
Compared to the material bound given by (16) in the main text, (\ref{eq:passivity_bounds}) essentially corresponds to a particular value of $p$. For example, given a lossless $\chi$ and narrow bandwidth $\Delta\omega \ll |\tilde{\omega}|$, we have 
\begin{equation*}
    f(\tilde{\omega}, \chi) \approx \frac{|\tilde{\omega}|}{\Delta\omega} \frac{\chi^2}{1+\chi}.
\end{equation*}
Now if we select a phase rotation $p$ such that $p_r=1$,
\begin{equation}
    p_i = \frac{\chi+1}{\chi} \frac{\Delta{\omega}}{\abs{\tilde{\omega}}},
\end{equation}
the material bound is
\begin{align*}
    \frac{|\tilde{\omega}|}{2} \Im{p/\chi^*}^{-1} \bra{\vb{E}_v}\ket{\vb{E}_v} = \frac{|\tilde{\omega}|}{2} \frac{|\tilde{\omega}|}{\Delta\omega} \frac{\chi^2}{1+\chi} \bra{\vb{E}_v}\ket{\vb{E}_v} = \frac{|\tilde{\omega}|}{2} f(\tilde{\omega}, \chi) \bra{\vb{E}_v}\ket{\vb{E}_v}
\end{align*}
which is exactly equivalent to the passivity bound. Checking (\ref{eq:cmax}) it is clear that with such a phase rotation $p$, $\Asym(p\U)$ remains positive definite.

\subsection{Bandwidth scaling of the passivity bounds}
The passivity bounds (\ref{eq:passivity_bounds}) have increased bandwidth scaling given an infinite halfspace design region coming from traveling wave contributions to $\bra{\vb{E}_v}\ket{\vb{E}_v}$. For example, given TM polarization we have
\begin{align*}
    \bra{\vb{E}_v^{TM}}\ket{\vb{E}_v^{TM}} &= \frac{\abs{\tilde{k}}^2}{8\pi} \int_{-\infty}^\infty \frac{ e^{-2 k_{yi} d}}{\abs{k_y}^2} \int_0^\infty e^{-2 k_{yi} y} \dd y \, \dd k_x \\
    &= \frac{\abs{\tilde{k}}^2}{8\pi} \int_{-\infty}^\infty \frac{ e^{-2 k_{yi}d} }{2\abs{k_y}^2 k_{yi}} \, \dd k_x.
\end{align*}
Now for traveling waves $\abs{k_x} < \tilde{k}_r$, it can easily be shown from $k_y = \sqrt{\tilde{k}^2 - k_x^2}$ that $k_{yi} \propto \tilde{k}_i$, leading to $\bra{\vb{E}_v^{TM}}\ket{\vb{E}_v^{TM}} \propto 1/\tilde{k}_i \propto 1/\Delta\omega$ (this analysis also applies to TE polarization). For lossless dielectrics, $f(\tilde{\omega}, \chi) \approx \frac{|\tilde{\omega}|}{\Delta\omega}$, giving the passivity bounds an overall $1/\Delta\omega^2$ scaling, as compared to the $1/\Delta\omega$ scaling for the full bounds. For lossy materials $f(\tilde{\omega}, \chi) \approx \frac{\abs{\chi}^2}{\chi_i}$, giving the passivity bounds an overal $1/\Delta\omega$ scaling, as compared to the $1/\Delta\omega^4$ scaling of the full bounds seen in Fig (3) of the main text. 

\bibliographystyle{plain}
\bibliography{LDOS}